\documentclass[reprint,superscriptaddress,aps,pra]{revtex4-1}
\usepackage{setspace}
\usepackage{amsmath}
\usepackage{amsfonts}
\usepackage{amssymb}
\usepackage{graphicx}
\usepackage{braket}
\usepackage{hyperref}
\usepackage[english]{babel}
\usepackage{geometry}

\newcommand{\Om}{\Omega_{m}}
\newcommand{\Gm}{\Gamma_{\!m}}
\newcommand{\Sm}{\Sigma_{m}}
\newcommand{\xzpf}{x_\mathrm{{zpf}}}
\newcommand{\oo}{\omega}
\newcommand{\oj}{\omega_{j}}
\newcommand{\opr}{\omega_\mathrm{probe}}
\newcommand{\octrl}{\omega_\mathrm{control}}

\newcommand{\keone}{\kappa_{e_{1}}}
\newcommand{\ketwo}{\kappa_{e_{2}}}
\newcommand{\Soo}{\Sigma_{\mathrm{o}_1}}
\newcommand{\Sot}{\Sigma_{\mathrm{o}_2}}

\newcommand{\aj}{a_{j}}
\newcommand{\daj}{\delta a_{j}}
\newcommand{\da}{\delta a}
\newcommand{\Djb}{\bar{\Delta}_{j}}
\newcommand{\Db}{\bar{\Delta}}

\newcommand{\dsp}{\delta s^{+}}
\newcommand{\dsm}{\delta s^{-}}

\newcommand{\C}{\mathcal{C}}

\newcommand{\T}{^{\text{T}}}
\newcommand{\dd}{\delta}
\newcommand{\as}{^\ast}
\newcommand{\dg}{^\dagger}
\newcommand{\im}{\mathrm{i}}
\begin{document}
\title{Nonreciprocity and magnetic-free isolation based on optomechanical interactions}
\author{Freek Ruesink}
\affiliation{Center for Nanophotonics, FOM Institute AMOLF, Science Park 104, 1098 XG Amsterdam, The Netherlands}
\author{Mohammad-Ali Miri}
\affiliation{Department of Electrical and Computer Engineering, The University of Texas at Austin, Austin, TX 78712, USA}
\author{Andrea Al\`{u}}
\affiliation{Department of Electrical and Computer Engineering, The University of Texas at Austin, Austin, TX 78712, USA}
\author{Ewold Verhagen}
\email{verhagen@amolf.nl}
\affiliation{Center for Nanophotonics, FOM Institute AMOLF, Science Park 104, 1098 XG Amsterdam, The Netherlands}
\date{\today}
\begin{abstract}
Photonic nonreciprocal components, such as isolators and circulators, provide highly desirable functionalities for optical circuitry.
This motivates the active investigation of mechanisms that break reciprocity, and pose alternatives to magneto-optic effects in on-chip systems. In this work, we use optomechanical interactions to strongly break reciprocity in a compact system.
We derive minimal requirements to create nonreciprocity in a wide class of systems that couple two optical modes to a mechanical mode, highlighting the importance of optically biasing the modes at a controlled phase difference.
We realize these principles in a silica microtoroid optomechanical resonator and use quantitative heterodyne spectroscopy to demonstrate up to 10 dB optical isolation at telecom wavelengths.
We show that nonreciprocal transmission is preserved for nondegenerate modes, and demonstrate nonreciprocal parametric amplification.
These results open a route to exploiting various nonreciprocal effects in optomechanical systems in different electromagnetic and mechanical frequency regimes, including optomechanical metamaterials with topologically non-trivial properties.
\end{abstract}

\maketitle
\onecolumngrid

Lorentz reciprocity stipulates that electromagnetic wave transmission is invariant under a switch of source and observer~\cite{Deak2012}, and its implications widely permeate physics.
To violate reciprocity and obtain asymmetric transmission, suitable forms of time-reversal symmetry breaking are required~\cite{Haldane2008}.
In optical and microwave systems this is usually achieved using magneto-optic material responses.
However, a vibrant search for alternative methods to break reciprocity, mimicking a magnetic bias, has taken shape in recent years~\cite{Sounas2013,Guo2015,Sayrin2015,Kim2015,Dong2015,Tzuang2014,Ranzani2015,Metelmann2015}.
This is fuelled by the typically weak magneto-optic coefficients in natural materials and/or their associated losses, and the technological promise of integrated on-chip nonreciprocal devices~\cite{Shoji2008}, including isolators and circulators.
A promising approach relies on spatiotemporal modulation of the refractive index to break time-reversal symmetry.
Such modulation allows imparting a nonreciprocal phase on the transfer of a signal between two optical modes~\cite{Fang2012a,Tzuang2014} or establishing a form of angular momentum biasing to create nonreciprocity~\cite{Sounas2013,Estep2014,Sounas2014}.

Pronounced optical time-modulation can be realized in cavity optomechanics~\cite{Aspelmeyer2014}, where a mechanical resonator's displacement $x$ alters the resonance frequency $\omega_{c}$ of an optical cavity.
Simultaneously, light can control the mechanical motion through radiation pressure, surpassing the need for external modulation. In recent years, this interaction's dynamics has been exploited for mechanical cooling, optical amplification, wavelength conversion and optomechanically induced transparency~\cite{Aspelmeyer2014,Massel2011,Hill2012,Lecocq2016,Weis2010} (OMIT).
Hafezi and Rabl~\cite{Hafezi2012} theoretically envisioned that optomechanical interactions in ring resonators can enable nonreciprocal responses, and associated asymmetric cavity spectra were recently observed~\cite{Kim2015,Dong2015,Shen2016}.
In other recent work, it was recognized that the mechanically-mediated signal transfer between two optical modes can be made nonreciprocal with suitable optical driving~\cite{Habraken2012,Xu2015}, a mechanism that enables phonon circulators and networks with topological phases for sound and light~\cite{Habraken2012,Peano2015,Schmidt2015}.
\begin{figure}[t]
\center
\includegraphics[width=0.85\linewidth]{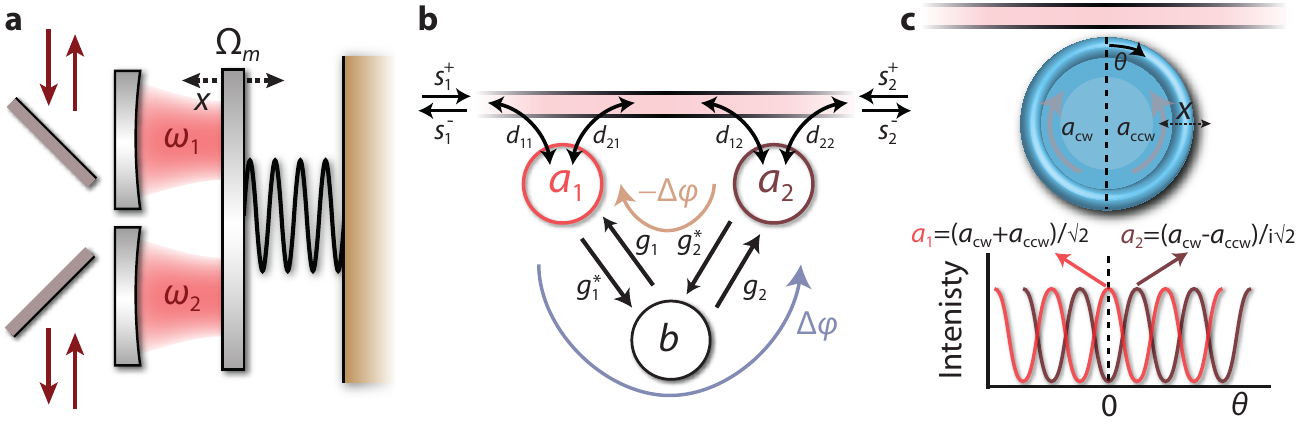}
\caption{\textbf{Nonreciprocity in a multimode optomechanical system.} \textbf{a)} Optomechanical system: two optical resonators at frequencies $\omega_{1}$ and $\omega_{2}$, both coupled to a mechanical mode at frequency $\Om$.
\textbf{b)} General description: the optical modes $(a_{1},a_{2})$ are coupled to a mechanical mode $b$ with coupling rates $g_{1}$ and $g_{2}$. 
The path $a_{1} \rightarrow b\rightarrow a_{2}$  picks up a phase $\Delta\phi=\arg(g_{2})-\arg(g_{1})$ that is opposite to that of the reversed path $a_{2} \rightarrow b \rightarrow a_{1}$. 
Two input/output ports ($s_1$ and $s_2$) are coupled to the optical modes with rates $d_{ij}$. Interfering both paths with direct scattering through the waveguide can build an optical isolator.
\textbf{c)} A ring-resonator supports even and odd optical modes $(a_{1},a_{2})$, superpositions of clockwise $(a_\mathrm{cw})$ and counter-clokwise $(a_{\mathrm{ccw}})$ propagating modes.  
As the two modes are $\pi/2$ out of phase with respect to a wave propagating in the waveguide, a control incident from a single input port fulfils the optimal driving conditions to break reciprocity. The graph sketches the spatial intensity profile of the two modes along the rim of the ring-resonator.}
\label{fig:optomechanics_NonRec}
\end{figure}

In the following, we show that all of the above systems can be understood within a general framework involving two optical modes coupled to a joint mechanical mode, which allows the derivation of general minimal conditions to achieve ideal optomechanical nonreciprocity.
Consider a basic system (Fig. ~\ref{fig:optomechanics_NonRec}a) of two optical modes with frequencies $(\omega_{1},\omega_{2})$, both coupled to a mechanical mode with frequency $\Om$.
The Hamiltonian of this system is~\cite{Aspelmeyer2014}
\begin{equation}
H=\hbar\sum_{j=1,2} \omega_{j} (x) \aj\dg\aj + \hbar\Om b\dg b,
\end{equation}
where $a$ and $b$ denote the photon and phonon annihilation operators, respectively, and $\oj (x) = \bar{\omega}_{j} - G_{j}x = \bar{\omega}_{j} - G_{j}\xzpf(b+b\dg)$, with $\xzpf$  the mechanical zero-point motion and $G_{j}$ the optical frequency shift per unit displacement. 
If both optical modes are driven by a strong coherent laser to an intracavity field $\alpha_{j} \exp(-\im\octrl t)$, the linearized Hamiltonian in a frame rotating at $\octrl$ reads
\begin{equation}
H_{L}=- \hbar\sum_{j=1,2} \Djb \daj\dg\daj + \hbar\Om b\dg b - 
\hbar \sum_{j=1,2} (g_{j}\as \daj b\dg +g_{j} \daj\dg b+g_{j}\as \daj b +g_{j}\daj\dg b\dg),
\label{eq:lin_OMhamiltonian}
\end{equation} 
where $\Djb = \octrl - \bar{\omega}_{j} + G_{j}\bar{x}$ is the control detuning from the cavity frequency (shifted by the mean displacement $\bar{x}$), and $\daj$ and $\daj\dg$ describe the small amplitude changes of the optical field. The interaction terms in the right describe coupling between the optical and mechanical modes at rates $g_{j}=G_{j}\xzpf\alpha_{j}$, controlled through the fields $\alpha_{j}$.

The crucial role of the relative phases of $g_j$ is immediately revealed when considering energy-conserving pairs that mediate inter-mode transfer. 
For example, photon annihilation in mode 1 upon phonon creation $(g_{1}\as\da_{1} b\dg)$, and the subsequent annihilation of the phonon with photon creation in mode 2 $(g_{2}\da_{2}\dg b)$ leads to a phase pickup $\Delta\phi=\arg(g_{2})-\arg(g_{1})$, whereas the reverse process provides an opposite phase $-\Delta\phi$~\cite{Habraken2012,Xu2015} (Fig.~\ref{fig:optomechanics_NonRec}b).
Strongest nonreciprocity is thus achieved when the two optical modes are driven with a phase difference~$=\pi/2$. 

Interestingly, this requirement is readily met in ring-resonators, such as the silica microtoroid studied here. 
This well-known optomechanical system supports a mechanical breathing mode coupled to an even and an odd optical mode (Fig. ~\ref{fig:optomechanics_NonRec}c)~\cite{Schliesser2008}. A control beam incident through an evanescently coupled waveguide excites an equal superposition of even and odd modes with $\pi/2$ phase difference~\cite{Yu2009}, such that the requirement on the control phase to maximally break reciprocity is automatically fulfilled. Note that our choice of the even/odd basis (in contrast to the clockwise/counterclockwise basis considered in other work~\cite{Hafezi2012,Kim2015,Shen2016}) immediately reveals the role of a nonreciprocal phase in intermode coupling, unifying the description of ring resonators and other systems. 

\begin{figure}[t]
\center
\includegraphics[width=0.75\linewidth]{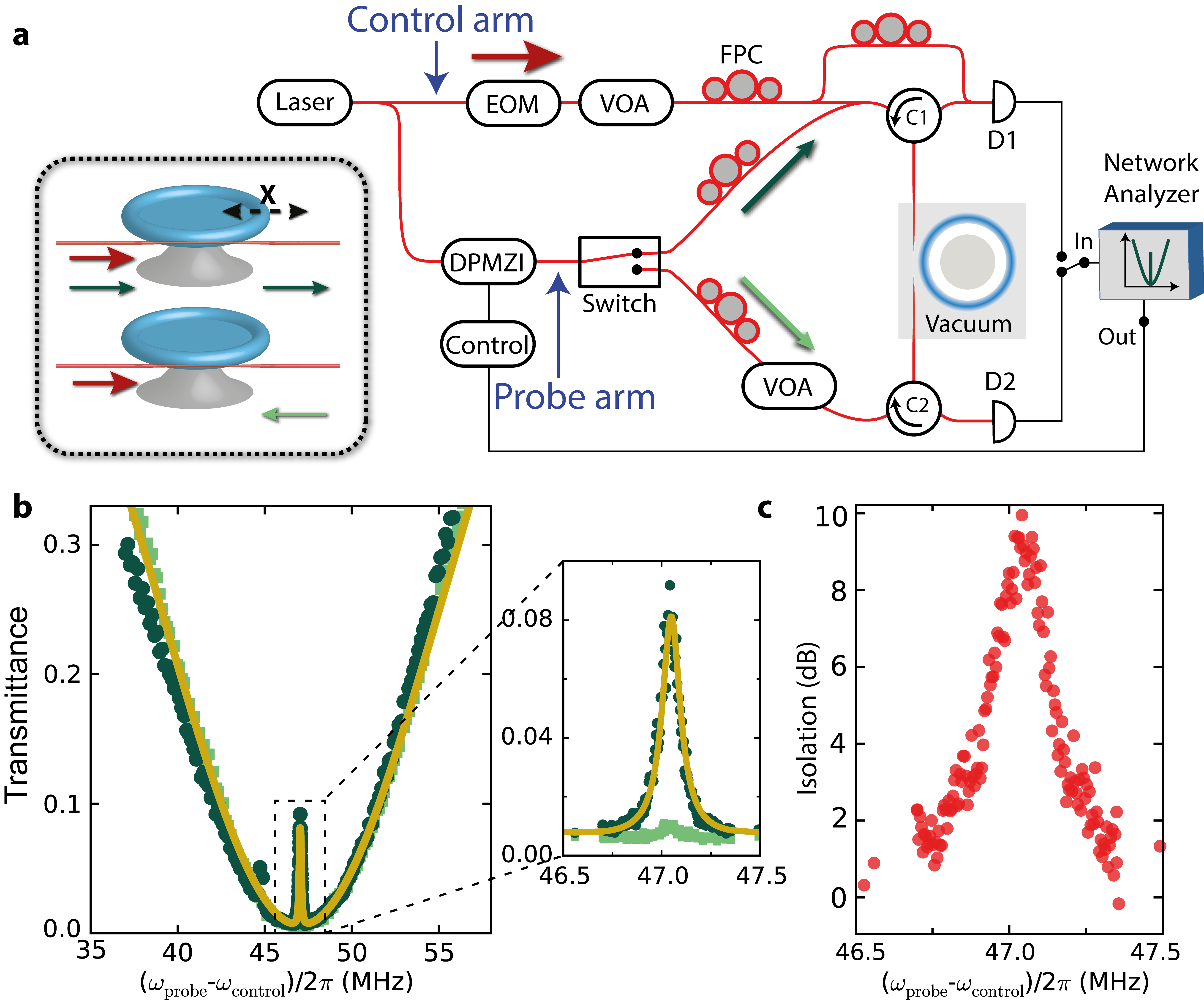}
\caption{
\textbf{Experimental setup and isolation.} 
\textbf{a)} A fibre-coupled laser signal is split into a control and probe arm, where the probe frequency is controlled through the output of a vector network analyser (VNA). An optical switch launches the probe into a tapered fibre either co-propagating or counter-propagating to the control beam. In both situations the transmittance at the probe frequency is extracted by analysing the beat with a calibrated control field using the VNA. The inset sketches the pump/probe configurations. 
\textbf{b)} Transmittance of the optical probe beam as a function of probe-control detuning with the control frequency ($\sim$17 $\mu$W) tuned to the red mechanical sideband. When the probe beam co-propagates (dark-green circles) with the control beam, an OMIT transmission window is observed, which is instead absent when the control and probe counter-propagate (light-green squares), resulting in non-reciprocal optical transmission. The solid yellow line is a fit of $|S_{21} |^{2}$ using independently determined values $(\Om,\Gm)/2\pi\approx$ (47.04 MHz, 87.7 kHz), yielding $(\kappa_{1,2},|g_{1}|)/2\pi=$ (28 MHz, 292 kHz)  and $\eta_{1,2}\approx0.45$.
\textbf{c)} Resulting isolation, quantified as the ratio of measured probe transmittance in both directions.}
\label{fig:optomechanics_isolation}
\end{figure}

The nature of the nonreciprocal response is determined by the direct coupling between the two channels: if it is forbidden (Fig. ~\ref{fig:optomechanics_NonRec}a), the system primarily functions as a nonreciprocal phase shifter. If a direct pathway exists (Fig. ~\ref{fig:optomechanics_NonRec}c), its interference with the resonant path that collects a nonreciprocal $\pi$ phase shift enables ideal isolation under appropriate conditions. 
In our experiment, we demonstrate optical isolation by studying the two-way transmittance of a probe signal at frequency $\opr$ through a tapered fibre that is coupled to a microcavity $(\omega_{1,2}/2\pi=$ 194.5 THz) with linewidth $\kappa/2\pi=$ 28 MHz. With the control laser incident from one direction, transmittance and optical isolation is quantified using a heterodyne spectroscopic technique, where a probe beam propagating in the forward or backward direction is recombined with the control, and their beat analyzed (see Fig.~\ref{fig:optomechanics_isolation}a and Supplementary Information).

The resulting probe transmittance (Fig.~\ref{fig:optomechanics_isolation}b) for $\bar{\Delta}_{1,2}=-\Om$ and near-critical coupling conditions shows a bidirectional transmission dip as the probe frequency is scanned across the cavity resonance. Importantly, the OMIT window~\cite{Weis2010}, which results from destructive intracavity interference of anti-Stokes scattering of the control beam from the probe-induced mechanical vibrations with the probe beam itself, is solely present for co-propagating control and probe (dark-green circles). For reversed probe direction OMIT is absent (light-green squares). The device thus acts as an optical isolator, reaching up to 10 dB of isolation (Fig.~\ref{fig:optomechanics_isolation}c).

To predict the magnitude of such nonreciprocal transmission, we use temporal coupled mode theory~\cite{Suh2004} to formulate the scattering matrix $S$ of a general system described by equation~(\ref{eq:lin_OMhamiltonian}), relating input $(\dsp_{j})$ and output $(\dsm_{j})$ waves at frequency $\opr$ in the ports $j=(1,2)$ via $(\dsm_{1},\dsm_{2})^{\text{T}}=S(\opr)(\dsp_{1},\dsp_{2})^{\text{T}}$. The dynamics of a two-mode system described by a linear time-evolution operator $\mathcal{M}$ reads 
\begin{equation}
\frac{d}{dt} \begin{pmatrix} \da_{1} \\ \da_{2}  \end{pmatrix} 
= \im\mathcal{M} \begin{pmatrix} \da_{1} \\ \da_{2} \end{pmatrix} 
+ D\T \begin{pmatrix} \dsp_{1}\\ \dsp_{2} \end{pmatrix},
\label{eq:timeEvolution}
\end{equation}
where $D$ describes the mutual coupling to the input/output fields. The output fields are found from
\begin{equation}
\begin{pmatrix} \dsm_{1} \\ \dsm_{2}  \end{pmatrix} 
= C \begin{pmatrix} \dsp_{1} \\ \dsp_{2} \end{pmatrix} 
+ D \begin{pmatrix} \da_{1}\\ \da_{2} \end{pmatrix},
\label{eq:outputFields}
\end{equation}
where $C$ describes the direct coupling between the two ports. Here we prescribe the individual optical modes to be reciprocal, such that coupling to in- and outgoing fields is identical~\cite{Suh2004}. In our system, it necessitates the choice of the even/odd mode basis. In the frequency domain, equations~(\ref{eq:timeEvolution},\ref{eq:outputFields}) yield the total scattering matrix 
\begin{equation}
S=C+\im D(M+\omega I)^{-1} D\T,
\label{eq:ScatMatrix_equation}
\end{equation}
with $I$ the identity matrix and $\omega=\opr-\octrl$.
In a general two-mode system, the difference between forward and backward complex transmission coefficients thus reads
\begin{equation}
S_{21}-S_{12} = \frac{\im\det(D)(m_{12}-m_{21})}{\det(M+\omega I)},
\label{eq:Sdif}
\end{equation}
showing that reciprocity can be broken as long as $\det(D)\neq 0$ and $m_{12}\neq m_{21}$ (with $m_{ij}$ the elements of $M$). This important result identifies the minimal conditions to break reciprocity; a full-rank $D$ matrix, requiring an asymmetry in the coupling between the two optical modes and the channels $s_1$ and $s_2$, and an asymmetric evolution matrix, enforcing the coupling rate from mode 1 to mode 2 to be different from that of mode 2 to mode 1.

The evolution matrix $M$ that describes optomechanical interactions (Fig.~\ref{fig:optomechanics_NonRec}) is obtained from the equations of motion 
\begin{equation}
\frac{d}{dt} \begin{pmatrix} \da_{1} \\ \da_{2}  \end{pmatrix} 
=\im \begin{pmatrix} \Db_1+\im\kappa_1/2 & 0 \\ 0 & \Db_2+\im\kappa_2/2 \end{pmatrix}
\begin{pmatrix} \da_{1} \\ \da_{2}  \end{pmatrix} 
+ \im \begin{pmatrix} g_1(b+b\dg) \\ g_2(b+b\dg)  \end{pmatrix} 
+ D^\text{T} \begin{pmatrix} \dsp_{1} \\ \dsp_{2}  \end{pmatrix} ,
\label{eq:EOM_optical}
\end{equation}
\begin{equation}
\frac{d}{dt}b = (-\im\Om-\Gm/2)\,b+\im(g_1\as\da_1+g_1\da_1\dg+g_2\as\da_2+g_2\da_2\dg),
\label{eq:EOM_mech}
\end{equation}
derived from the linearized Hamiltonian (\ref{eq:lin_OMhamiltonian}) including dissipation.
Note that we have set coupling between the optical modes to zero, which can always be realized through diagonalization (see Supplementary Information). 
Solving these equations in the frequency domain, applying the rotating wave approximation and using the input-output relation~(\ref{eq:outputFields}), the evolution matrix ($M+\omega I$) for $\omega\approx\pm\Om$ reads 
\begin{equation}
M+\omega I =
\begin{pmatrix}
\Soo \mp \frac{|g_1|^2}{\Sm^\pm} & \mp \frac{g_1 g_2\as}{\Sm^\pm} \\
\mp \frac{g_1\as g_2}{\Sm^\pm} & \Sot \mp \frac{|g_2|^2}{\Sm^\pm}
\end{pmatrix}.
\label{eq:matrixM}
\end{equation}
Here, $\Sigma_{\mathrm{o}_j} = \omega+\Db_j+\im\kappa_j/2$ is the inverse optical susceptibility, $\Sm^\pm =\omega\mp\Om+\im\Gm/2$ the inverse mechanical susceptibility and $\Gm$ the mechanical damping rate. Importantly, $(m_{12}-m_{21}) \propto \sin\Delta\phi$, highlighting the importance of the control phase difference to obtain nonreciprocal transmission. 

By combining~(\ref{eq:Sdif}) and~(\ref{eq:matrixM}), and introducing the cooperativities $\C_j=4|g_j|^2/(\kappa_j\Gm)$, the asymmetric transmission through a two-mode system can be written (Supplementary Information) as 
\begin{equation}
S_{21}-S_{12} = 2\sin\Delta\phi  
\sqrt{\eta_1\eta_2}\frac{\mp 2\sqrt{\C_1\C_2}}{(\dd_\pm+\im)(\dd_1+\im)(\dd_2+\im)\mp(\C_2(\dd_1+\im)+\C_1(\dd_2+\im))},
\label{eq:SdifC}
\end{equation}
where we defined the relative detuning of the probe frequency $\dd_\pm\equiv(\omega\mp\Om)/(\Gm/2)$ and $\dd_j\equiv(\omega+\Db_j)/(\kappa_{j}/2)$ from mechanical and optical resonance, and $\eta_j$ is the fraction of energy mode $j$ radiates in both output channels. 
Inspection of~(\ref{eq:SdifC}) shows that the magnitude of asymmetric transmission at critical coupling $(\eta_{1,2}=1/2)$ is maximally 1, when the cooperativities are large and equal. 
These conditions, implemented in our experiment, enable the observed strong photonic isolation.
\begin{figure}[t]
\center
\includegraphics[width=0.75\linewidth]{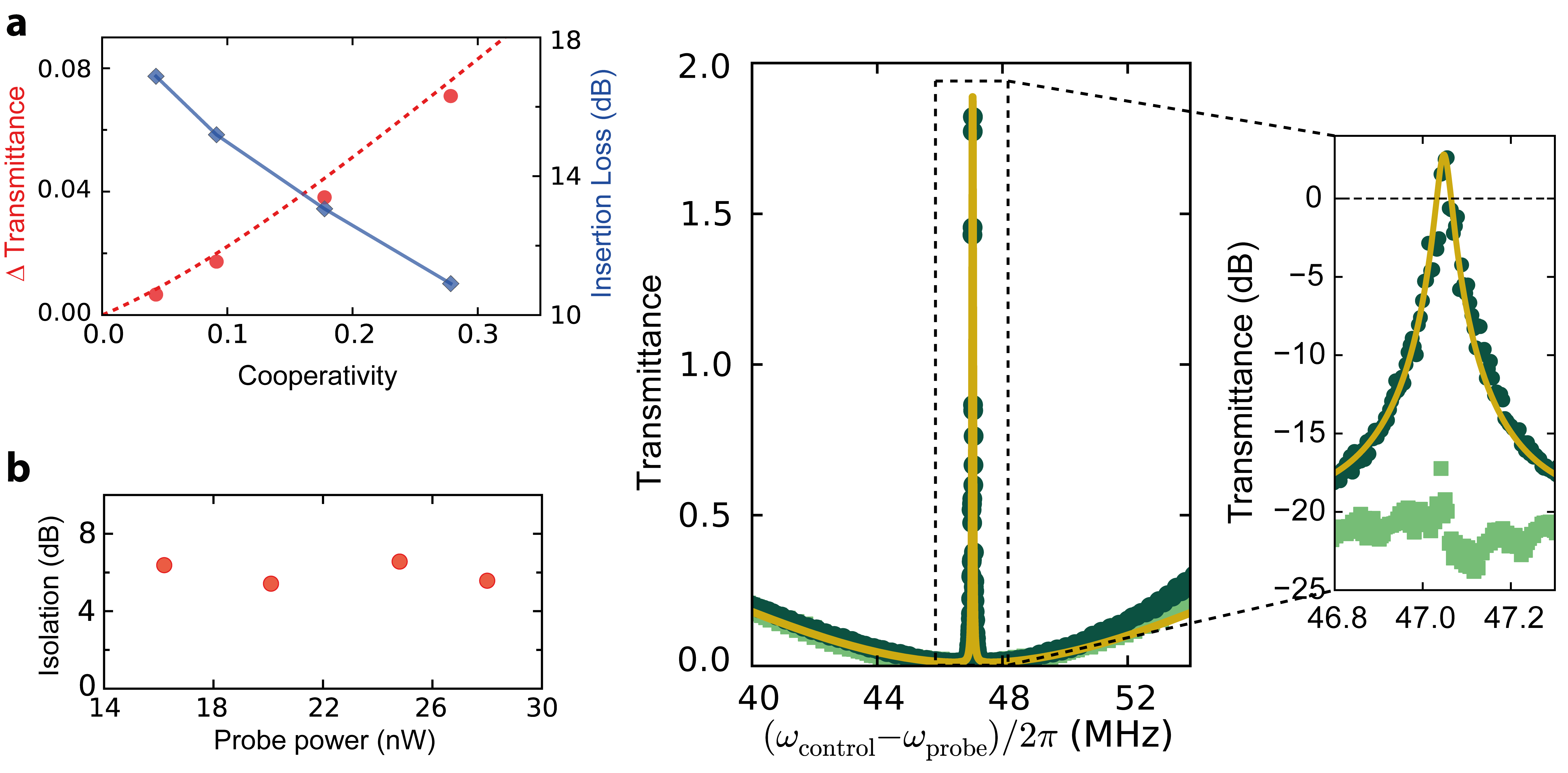}
\caption{
\textbf{Power-dependence and nonreciprocal amplification.} 
\textbf{a)} Difference between forward/backward transmissivities (measured, red circles, and theory, dashed line) with respect to cooperativity, directly proportional to the control laser power. Together with an increase in contrast, the insertion losses (blue diamonds) decrease with increasing cooperativity
\textbf{b)} The isolation as a function of probe power sent through the fibre. The physical mechanism behind optical isolation is linear, and thus does not depend on probe power.
\textbf{c)} When the control beam is tuned to the blue-sideband of the cavity, it can parametrically amplify the probe beam that co-propagates with it through the fibre (dark-green circles). In contrast, the counter-propagating probe beam (light-green squares) experiences normal cavity extinction, thus yielding a non-reciprocal amplifier. With amplification of $\sim 3$dB, the non-reciprocal difference in transmission is approximately 23 dB. The solid yellow line is a fit of $|S_{21}|^{2}$ yielding $(\kappa_{1,2},|g_1|)/2\pi\approx$(28 MHz, 454 kHz) and $\eta_{1,2}\approx0.46$.
}
\label{fig:powerDep_Amp}
\end{figure}

For degenerate optical modes and the control field tuned to either mechanical sideband, the maximum contrast between forward and backward transmittance is $\Delta T=|S_{21}|^{2}-|S_{12}|^2 \approx (\C^{-1}\pm 1)^{-2}$ at $(\omega=\pm\Om,\Db_{1,2}=\mp\Om)$, where $\C=2\C_1=2\C_2$. The pronounced increase of $\Delta T$ with increasing cooperativity, and concomitant decrease of insertion loss, are confirmed by varying the optical drive power (Fig.~\ref{fig:powerDep_Amp}a). The mechanism has strong potential for near-ideal isolation at negligible insertion losses, for example in optimized silica microtoroids, where $\C\approx 500$ was demonstrated~\cite{Verhagen2012}. Moreover, cooperativity enhances the bandwidth, which is ultimately limited by the optical linewidth~\cite{Hafezi2012}. An important aspect of this mechanism is that the isolation is independent of probe power (Fig.~\ref{fig:powerDep_Amp}b), differing fundamentally from mechanisms exploiting static nonlinearity~\cite{Manipatruni2009,Fan2012} to create asymmetric transmission.

For blue-detuned control $(\Db_{1,2}=+\Om)$, the probe beam experiences parametric amplification if control and probe are co-propagating, while it is fully dissipated when counter-propagating with the pump, thus yielding a nonreciprocal optical amplifier  (Fig.~\ref{fig:powerDep_Amp}c). This feature could pose interesting signal processing functionality, including nonreciprocal narrowband RF filtering and insertion loss compensation.

Note that for the even/odd optical modes in the evanescently coupled ring-resonator system that is experimentally studied here, where $c_{ij}=1$, the $D$ matrix is constrained~(\cite{Suh2004}, Supplementary Information) to 
\begin{equation}
D=\frac{1}{\sqrt{2}} 
\begin{pmatrix}
\im\sqrt{\kappa_{e_{1}}} & -\sqrt{\kappa_{e_{2}}} \\
\im\sqrt{\kappa_{e_{1}}} & \sqrt{\kappa_{e_{2}}} 
\end{pmatrix}.
\label{eq:Dmatrix}
\end{equation}
Together with~(\ref{eq:ScatMatrix_equation}) and~(\ref{eq:matrixM}), this $D$ matrix yields the expressions for the scattering matrix elements (Supplementary Information) used to fit the data in figures 2-4.

\begin{figure}[t]
\center
\includegraphics[width=0.7\linewidth]{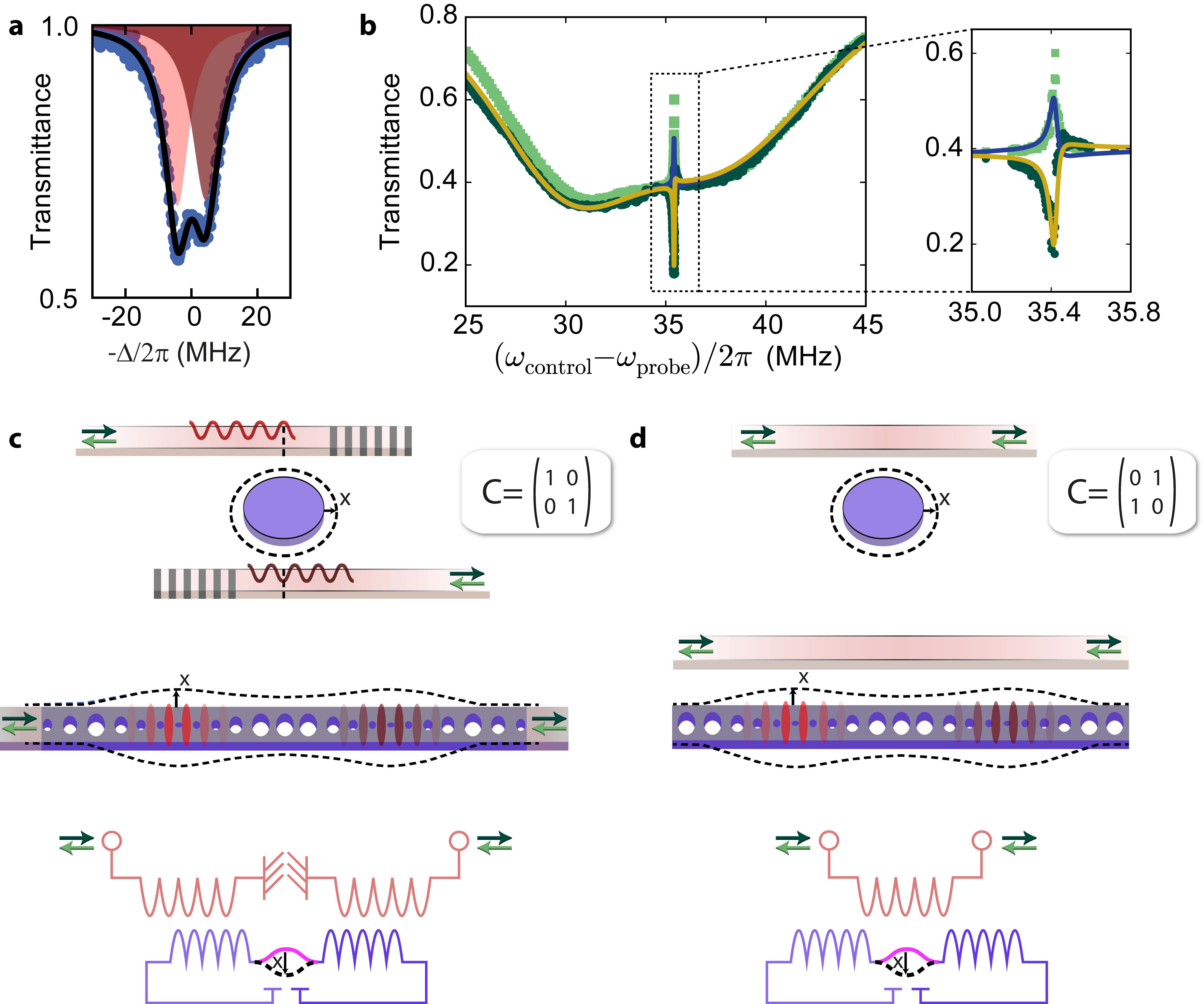}
\caption{
\textbf{Mode-splitting and general optomechanical nonreciprocity.} 
\textbf{a)} Transmittance of an optical split-mode (splitting $\sim$8.6 MHz) as a function of laser-cavity detuning, obtained by sweeping the laser frequency and measuring the resulting transmittance using an oscilloscope. The horizontal axis is calibrated using the EOM placed in the control arm (see Fig~\ref{fig:optomechanics_isolation}a).  
\textbf{b)} Transmittance of the optical probe beam as a function of control-probe detuning with the control frequency fixed at the blue mechanical sideband. When the probe beam co-propagates (dark-green circles) with the control beam an optomechanically induced absorption window appears, while the oppositely propagating probe (light-green squares) experiences increased transmission. The solid lines represent a combined fit of $|S_{21}|^2$ (yellow) and $|S_{12}|^2$ (blue) to the dark-green circles and light-green squares, respectively. 
\textbf{c)/d)} Different optomechanical systems that can yield non-reciprocal behaviour.
\textbf{c)} If the optical and mechanical resonators are placed in the direct propagation path, the displayed systems have a diagonal direct scattering matrix $C$, which builds a non-reciprocal phase shifter.
\textbf{d)} However, if the optomechanical system provides an extra (non-reciprocal) propagation path in addition to the direct scattering path, the $C$ matrix has off-diagonal elements which can interfere with the non-reciprocal path to yield isolation or amplification.   
}
\label{fig:splitting}
\end{figure}

Importantly, equation~(\ref{eq:SdifC}) shows that strong nonreciprocity can also be obtained without optical degeneracy. If the two modes have different frequency and/or linewidth, an optimal control frequency can be chosen to satisfy $\dd_1=-\dd_2=\beta$. Then asymmetric transmittance is maximally
\begin{equation}
\Delta T=\frac{\C(\C\pm 2\beta^{2})}{(1\pm\C+\beta^{2})^{2}},
\label{eq:DeltaT}
\end{equation}
showing that larger cooperativity can compensate the effects of mode splitting for $\beta>1$.
Figures~\ref{fig:splitting}a-b show nonreciprocal transmission with a split optical mode. With a blue-detuned control and $\eta_{1,2}<1/2$, we observe respectively optomechanically induced transparency and absorption in opposite directions, as confirmed by theory. Since here $\beta\ll\Om/\kappa_{1,2}$, the control beam incident from one side still ensures $\Delta\phi\approx\pi/2$ and $\C_1\approx\C_2$, resulting in large contrast. In a more general case, optimal conditions may be implemented, for example by supplying control fields with suitable phase and amplitude through both input waveguides. Importantly, the fact that nonreciprocity can be obtained without optical degeneracy increases the range of systems that may be employed.
%

We stress that the demonstrated principles are not limited to the experimental implementation shown here, but can be realized in a wide range of optomechanical platforms~\cite{Aspelmeyer2014}, including photonic crystal resonators~\cite{Hill2012}, LC-circuits~\cite{Lecocq2016}, and various ring-resonator systems (Fig.~\ref{fig:splitting}c-d). The specific nonreciprocal functionality is governed by the way these systems are coupled to input/output channels. This, in turn, is directly related to the nonresonant scattering matrix $C$, as reciprocity of optical modes dictates $CD^*=-D$~\cite{Suh2004}. For the scenarios in figure~\ref{fig:splitting}c, described by a diagonal $C$ matrix, each waveguide couples to a single optical mode, and the system  operates as a nonreciprocal phase shifter. In contrast, isolation is achieved if $C$ is the exchange matrix, meaning a direct path between the two ports is present (Fig.~\ref{fig:splitting}d). With suitable cooling at high cooperativity, the functionality can be extended to the quantum domain at negligible added noise~\cite{Hafezi2012}. We note that nonreciprocity occurs also outside the resolved sideband regime, although the behaviour there is more complex due to mixing of sidebands at $\pm\omega$.

In conclusion, we demonstrated and quantified nonreciprocal transmission through a compact optomechanical isolator and parametric amplifier, and developed a general theory explaining the mechanism and unifying the description of various implementations of optomechanical nonreciprocity in multimode systems. Our findings identify two general requirements for any optomechanical system to optimally break reciprocity: asymmetric coupling of the optical modes to input/output channels, and their drive with a $\pi/2$ phase difference. Extending the demonstrated principles to more modes or channels would enable a variety of nonreciprocal functionality for both light and sound, including on-chip circulation, gyration~\cite{Habraken2012} and enhanced isolation bandwidth. Finally, these nonreciprocal systems can form the unit cell of optomechanical metamaterials with topologically non-trivial properties, where the nonreciprocal phase takes the role of an effective gauge field to establish new phases for sound and light~\cite{Peano2015,Schmidt2015}.\\

This work is part of the research programme of the Foundation for Fundamental Research on Matter (FOM), which is part of the Netherlands Organisation for Scientific Research (NWO). It has been  supported by the Office of Naval Research, and the Simons Foundation.\\
 

%


\renewcommand{\theequation}{S\arabic{equation}}
\renewcommand{\thesection}{\Alph{section}}

\setcounter{equation}{0}
\newpage
\large
\textbf{Supplementary Information for "Nonreciprocity and magnetic-free isolation based on optomechanical interactions"}

\small
\section{General requirements of non-reciprocity in a two-mode optical system}
In general, the temporal-evolution equations of a two-mode system can be described via [1]:
\begin{equation}
\frac{d}{dt} \begin{pmatrix} \da_{1} \\ \da_{2}  \end{pmatrix} 
= \im\mathcal{M} \begin{pmatrix} \da_{1} \\ \da_{2} \end{pmatrix} 
+ D^{\text{T}} \begin{pmatrix} \dsp_{1}\\ \dsp_{2} \end{pmatrix},
\label{eq:SI_timeEvolution}
\end{equation}
\begin{equation}
\begin{pmatrix} \dsm_{1} \\ \dsm_{2}  \end{pmatrix} 
= C \begin{pmatrix} \dsp_{1} \\ \dsp_{2} \end{pmatrix} 
+ D \begin{pmatrix} \da_{1}\\ \da_{2} \end{pmatrix},
\label{eq:SIoutputFields}
\end{equation}
where $\mathcal{M}$ is the linear evolution operator for the two optical modes, the matrix $D$ describes the coupling between the two ports and optical modes and $C$ describes the direct scattering path between the two ports. Note that $\dsp_{j}$ and $\dsm_{j}$ are normalized such that $|\dsp_j|^2$ is the input photon flux in channel $j$. In writing these equations, we have enforced the optical modes to each satisfy reciprocity, in the sense that the in-and out-coupling rates of a particular mode and port are equal, thus the coupling is described through the same matrix $D$ (and its transpose) in~(\ref{eq:SI_timeEvolution}) and~(\ref{eq:SIoutputFields}). In the frequency domain $\left((a[\omega],s[\omega])=\int(a(t),s(t)) \exp(\im\omega t)\, d t \right)$, equation~(\ref{eq:SI_timeEvolution}) can be written as
\begin{equation}
-\im(M+\oo I) \begin{pmatrix} \da_{1} \\ \da_{2} \end{pmatrix}
= D\T \begin{pmatrix}
\dsp_1 \\ \dsp_2
\end{pmatrix},
\end{equation}
where $I$ represents the 2$\times$2 identity matrix. The scattering matrix, defined as 
\begin{equation}
\begin{pmatrix} \dsm_1 \\ \dsm_2  \end{pmatrix}
= S \begin{pmatrix} \dsp_1 \\ \dsp_2 \end{pmatrix},
\end{equation}	
is now directly obtained to be 
\begin{equation}
S = C + \im D(M+\oo I)^{-1}D\T .
\label{eq:SI_GenScatMat}
\end{equation}
A direct calculation of the off-diagonal scattering elements (forward and backward transmission coefficients) reveals that
\begin{equation}
S_{12}=c_{12}+\im\frac{(m_{11}+\oo)d_{12}d_{22}-m_{12}d_{11}d_{22}-m_{21}d_{12}d_{21}+(m_{22}+\oo)d_{11}d_{21}}{\det(M+\oo I)},
\end{equation}
\begin{equation}
S_{21}=c_{21}+\im\frac{(m_{11}+\oo)d_{12}d_{22}-m_{12}d_{12}d_{21}-m_{21}d_{11}d_{22}+(m_{22}+\oo)d_{11}d_{21}}{\det(M+\oo I)}.
\end{equation}
The contrast between forward and backward transmission coefficients now reads
\begin{equation}
S_{21}-S_{12} = \frac{\im\det(D)(m_{12}-m_{21})}{\det(M+\omega I)}.
\label{eq:SI_Sdif}
\end{equation}
From this formula, it becomes immediately clear that the necessary condition for breaking reciprocity is to ensure $\det (D)\times (m_{12}-m_{21})\neq 0$, which requires: (a) $\det (D)\neq 0$, and (b) $m_{12}\neq m_{21}$. As discussed in the main text, the first condition simply requires $D$ to be a full rank matrix, which is expected as in the case of a non-full-rank $D$ both optical modes are symmetrically coupled to the two ports, thus preserving  reciprocity.

\section{Breaking the symmetry of the evolution matrix through optomechanical coupling}
Here we consider the optical modes to have generally different resonant frequencies $\oo_{1,2}$ and energy decay rates $\kappa_1$ and $\kappa_2$. In addition, we describe the mechanical mode by its resonance frequency $\Om$ and loss rate $\Gm$. Starting from the linearized Hamiltonian in equation~(\ref{eq:lin_OMhamiltonian}) of the main text, the equations of motion for the photon and phonon annihilation operators can be written as:
\begin{equation}
\frac{d}{dt} \begin{pmatrix} \da_{1} \\ \da_{2}  \end{pmatrix} 
=\im \Theta \begin{pmatrix} \da_{1} \\ \da_{2}  \end{pmatrix} 
+ \im \begin{pmatrix} g_1(b+b\dg) \\ g_2(b+b\dg)  \end{pmatrix} 
+ D\T \begin{pmatrix} \dsp_{1} \\ \dsp_{2}  \end{pmatrix} ,
\label{eq:SI_EOM_optical}
\end{equation}
\begin{equation}
\frac{d}{dt}b = (-\im\Om-\Gm/2)\,b+\im(g_1\as\da_1+g_1\da_1\dg+g_2\as\da_2+g_2\da_2\dg),
\label{eq:SI_EOM_mech}
\end{equation}
where in this relation $g_{1,2}$ represent the enhanced optomchanical coupling rates defined as $g_1=G_1\xzpf\alpha_1$ and $g_2=G_2\xzpf\alpha_2$, while $\Theta$ is a 2$\times$2 matrix which describes the evolution of the optical modes in the absence of the optomechanical interactions and in the absence of port excitations. 
In general $\Theta$ can be written as $\Theta=\Omega +\im K$ where $\Omega$ represents the resonance frequencies and the mutual coupling of the two modes on diagonal and off-diagonal elements  respectively. 
The matrix $K=K_0+K_e$, on the other hand, represents the total losses of the two modes $\kappa_{1,2}=\kappa_{0_{1,2}}+\kappa_{e_{1,2}}$ which includes the intrinsic losses $(\kappa_{0_{1,2}})$ due to absorption and scattering as well as the out-coupling losses $(\kappa_{e_{1,2}})$ due to the leakage of the optical modes into the two ports.
Even though in general $\Theta$ is not diagonal, it is always possible to choose a proper eigenmode basis such that $\Theta$ becomes diagonal (see section C). In that case, it can be written as
\begin{equation}
\Theta=\begin{pmatrix}
\Db_1 + \im\frac{\kappa_1}{2} & 0 \\
0 & \Db_2 + \im\frac{\kappa_2}{2}
\end{pmatrix},
\label{eq:SI_Theta}
\end{equation}
where $\Db_1=\octrl-\bar{\oo}_1+G_1\bar{x}$, $\Db_2=\octrl-\bar{\oo}_2+G_2\bar{x}$ represent the modified frequency detunings of the two optical modes from the driving lasers. In the frequency domain, equations~(\ref{eq:SI_EOM_optical}) and~(\ref{eq:SI_EOM_mech}) become 
\begin{equation}
\im \begin{pmatrix} \Soo & 0 \\ 0 & \Sot  \end{pmatrix} 
\begin{pmatrix} \da_{1} \\ \da_{2}  \end{pmatrix} 
+ \im \begin{pmatrix} g_1(b+b\dg) \\ g_2(b+b\dg)  \end{pmatrix} 
+ D\T \begin{pmatrix} \dsp_{1} \\ \dsp_{2}  \end{pmatrix} =0,
\label{eq:SI_EOM_FT_optical}
\end{equation}
\begin{equation}
\im\Sm^+ \,b  +\im(g_1\as\da_1+g_1\da_1\dg+g_2\as\da_2+g_2\da_2\dg) = 0 ,
\label{eq:SI_EOM_FT_mech}
\end{equation}
where we introduce a shorthand notation for the Fourier transformed creation operators $\da\dg = (\da\dg)[\oo] = (\da[-\oo])\dg$, and defined the inverse susceptibilities
\begin{equation}
\Sm^\pm\equiv\oo\mp\Om+\im\frac{\Gm}{2}; \;\;\;\; \Sigma_{\mathrm{o}_{1,2}}\equiv\oo+\Db_{1,2}+\im\frac{\kappa_{1,2}}{2}
\end{equation}
Using~(\ref{eq:SI_EOM_FT_optical}) and~(\ref{eq:SI_EOM_FT_mech}) along with its Hermitian conjugate, it is straightforward to show that
\begin{align}
\im \begin{pmatrix} \Soo & 0 \\ 0 & \Sot  \end{pmatrix} 
\begin{pmatrix} \da_{1} \\ \da_{2}  \end{pmatrix}
+&\im \left(\frac{1}{\Sm^-}-\frac{1}{\Sm^+}\right) 
\begin{pmatrix} |g_1|^2 & g_1 g_2\as \\ g_1\as g_2 & |g_2|^2 \end{pmatrix}
\begin{pmatrix} \da_{1} \\ \da_{2}  \end{pmatrix} \nonumber \\
+&\im \left(\frac{1}{\Sm^-}-\frac{1}{\Sm^+}\right) 
\begin{pmatrix} g_1^2 & g_1 g_2 \\ g_1 g_2 & g_2^2 \end{pmatrix}
\begin{pmatrix} \da_{1}\dg \\ \da_{2}\dg  \end{pmatrix}
+ D\T \begin{pmatrix} \dsp_{1} \\ \dsp_{2}  \end{pmatrix} =0.
\label{eq:SI_nonRSB}
\end{align}
For the remainder we will consider operation in the resolved side-band regime, which allows ignoring the non-resonant terms involving $a_{1,2}\dg$ in equation~(\ref{eq:SI_nonRSB}), such that it simplifies to 
\begin{equation}
\im \left[ \begin{pmatrix} \Soo & 0 \\ 0 & \Sot  \end{pmatrix} 
+\im \left(\frac{1}{\Sm^-}-\frac{1}{\Sm^+}\right) 
\begin{pmatrix} |g_1|^2 & g_1 g_2\as \\ g_1\as g_2 & |g_2|^2 \end{pmatrix} \right]
\begin{pmatrix} \da_{1} \\ \da_{2}  \end{pmatrix}
+ D\T \begin{pmatrix} \dsp_{1} \\ \dsp_{2}  \end{pmatrix} =0. 
\label{eq:SI_RSB}
\end{equation}
Given that we are particularly interested in probe signals that are detuned from the control by approximately the mechanical resonance frequency $(|\oo\mp\Om|\ll\Om)$, one can always neglect one of the two terms involving mechanical susceptibilities $1/\Sm^\pm$. For $\Db_j\approx\mp\Om$, the frequency-domain evolution matrix $M$ can now be found from equation~(\ref{eq:SI_RSB}) to be 
\begin{equation}
M=\begin{pmatrix} 
\Db_1 + \im\frac{\kappa_1}{2} & 0 \\
0 & \Db_2 + \im\frac{\kappa_2}{2}
\end{pmatrix} 
\mp \frac{1}{\Sm^\pm} 
\begin{pmatrix} |g_1|^2 & g_1 g_2\as \\ g_1\as g_2 & |g_2|^2 \end{pmatrix}.
\end{equation}
Clearly, in order to break the symmetry of $M$, which is a necessary condition for breaking the reciprocity of the system, one needs to enforce $g_1 g_2\as \neq g_1\as g_2$, thus requiring $\Delta\phi=\arg\left(g_2\right)-\arg\left(g_{1} \right)\neq n \pi$ where $n \in\mathbb{N}$. The resulting scattering matrix reads
\begin{equation}
S=C+\im D(M+\omega I)^{-1} D^\text{T} =C+\im D
\begin{pmatrix}
\Soo \mp \frac{|g_1|^2}{\Sm^\pm} & \mp \frac{g_1 g_2\as}{\Sm^\pm} \\
\mp \frac{g_1\as g_2}{\Sm^\pm} & \Sot \mp \frac{|g_2|^2}{\Sm^\pm}
\end{pmatrix}^{\!\!-1}
D^\text{T}.
\label{eq:SI_Smatrix}
\end{equation}
Using equation~(\ref{eq:SI_Sdif}), the difference in transmission $S_{21}-S_{12}$ is directly given as
\begin{equation}
S_{21}-S_{12} = 2\sin\Delta\phi \frac{\det D}{\sqrt{\kappa_1\kappa_2}} \frac{\mp 2\sqrt{\C_1\C_2}}{(\dd_\pm+\im)(\dd_1+\im)(\dd_2+\im)\mp(\C_2(\dd_1+\im)+\C_1(\dd_2+\im))},
\label{eq:SI_SdifC}
\end{equation}
where again the upper and lower signs relate to red $(\Db_{1,2}\approx -\Om)$ and blue $(\Db_{1,2}\approx +\Om)$  detuned regimes respectively, and we have defined 
\begin{equation}
\C_j\equiv\frac{4|g_j|^2}{\kappa_j\Gm}, \,\,\, \dd_\pm\equiv\frac{\oo\mp\Om}{\Gm/2}, \,\,\, 
\dd_j\equiv\frac{\oo+\Db_j}{\kappa_j/2}.
\end{equation}
Alternatively, one can obtain $S_{12}$ and $S_{21}$ separately, which are given by:
\begin{equation}
S_{12} = c_{12}+\im\frac{2A\pm 2\sqrt{\C_1\C_2}(d_{12}d_{21}e^{\im\Delta\phi}+d_{11}d_{22}e^{-\im\Delta\phi}}{\sqrt{\kappa_1\kappa_2}[(\dd_\pm+\im)(\dd_1+\im)(\dd_2+\im)\mp(\C_2(\dd_1+\im)+\C_1(\dd_2+\im))]},
\label{eq:SI_S12}
\end{equation}
\begin{equation}
S_{21} = c_{21}+\im\frac{2A\pm 2\sqrt{\C_1\C_2}(d_{11}d_{22}e^{\im\Delta\phi}+d_{12}d_{21}e^{-\im\Delta\phi}}{\sqrt{\kappa_1\kappa_2}[(\dd_\pm+\im)(\dd_1+\im)(\dd_2+\im)\mp(\C_2(\dd_1+\im)+\C_1(\dd_2+\im))]},
\label{eq:SI_S21}
\end{equation}
where 
\begin{equation}
A=d_{11}d_{21}\sqrt{\frac{\kappa_2}{\kappa_1}}[(\dd_\pm+\im)(\dd_2+\im)\mp\C_2]+d_{12}d_{22}\sqrt{\frac{\kappa_1}{\kappa_2}}[(\dd_\pm+\im)(\dd_1+\im)\mp\C_1],
\end{equation}
Note that these expressions are general in the sense that they are valid regardless of the way the modes are coupled to the ports. In section E we will consider a specific implementation. 

\section{Diagonalization of the optical evolution matrix}
Here we show that in general it is always possible to find a proper eigenbasis that diagonalizes the evolution matrix of an optical system. To show this, consider an optical system described through the coupled mode equations:
\begin{equation}
\frac{d}{dt} \mathbf{a} = \im\Theta \mathbf{a} + D\T \mathbf{s_+},
\label{eq:SI_modeEvo}
\end{equation}
\begin{equation}
\mathbf{s_-} = C \mathbf{s_+} + D \mathbf{a},
\label{eq:SI_inputOutput}
\end{equation}
where $\mathbf{a}=(a_1\;a_2)\T$ represents the modal amplitude of the two modes and $\mathbf{s_\pm}=(s_1^\pm \; s_2^\pm)\T$ represents the vector of inputs/outputs at two ports. The evolution matrix $\Theta$ can in general be written in form of:
\begin{equation}
\Theta = \begin{pmatrix} \Delta_1 & \mu \\ \mu & \Delta_2 \end{pmatrix} 
+\im/2 \begin{pmatrix} \kappa_1 & \kappa_r \\ \kappa_r & \kappa_2 \end{pmatrix}
\end{equation}
where in this relation $\Delta_{1,2}$ represent the frequency detunings of the two modes with respect to a central resonance frequency, $\mu$ represents the mutual coupling and $\kappa_1, \kappa_2$ and $\kappa_r$ are the optical losses due to intrinsic losses and leakage to ports as well as mutual coupling. Defining $X_R$ and $X_L$ as two matrices with the right and left eigenvectors of $\Theta$ as their columns and rows respectively, one can write [2]:
\begin{align}
\Theta X_R &= X_R \Lambda, \label{eq:SI_diag1}\\
X_L \Theta &= \Lambda X_L,\label{eq:SI_diag2}
\end{align}
where $\Lambda$ is the diagonal matrix of eigenvalues. Given that $\Theta$ is symmetric, the left and right eigenvectors are related via $X_L=X_R\T$. On the other hand, one can show that $X_L X_R=X_R X_L$ is diagonal.  
Therefore through a proper normalization of the eigenvectors one can write $X_R\T X_R=X_R X_R\T=I$, thus $X_R$ is an orthogonal matrix. Equations~(\ref{eq:SI_modeEvo}) and~(\ref{eq:SI_inputOutput}) can now be written as:
\begin{align}
\frac{d}{dt} X_R \,\mathbf{a}&=\im X_R \Theta X_R\T X_R \,\mathbf{a} + X_R D\T \mathbf{s_+},\\
\mathbf{s_-} &= C\mathbf{s_+} + D X_R\T X_R \, \mathbf{a},
\end{align}
which can be written in form of equations~(\ref{eq:SI_modeEvo}) and~(\ref{eq:SI_inputOutput}) as 
\begin{align}
\frac{d}{dt} \,\mathbf{a'}&=\im \Theta' \,\mathbf{a'} + D^{'\text{T}} \mathbf{s_+},\\
\mathbf{s_-} &= C\mathbf{s_+} + D' \, \mathbf{a'},
\end{align}
Here, $\mathbf{a'}=X_R \,\mathbf{a}$ is the new basis and $D'=D X_R\T$ represents the transformed mode-port coupling matrix. In the new eigenmode basis, the evolution matrix $\Theta'$ is defined as:
\begin{equation}
\Theta'=X_R \Theta X_R\T.
\end{equation}
Given the fact that $\Theta$ is symmetric, equations~(\ref{eq:SI_diag1}) and~(\ref{eq:SI_diag2})  directly imply that      $\Theta'$ is diagonal. Therefore, without loss of generality one can consider the optical evolution matrix as follows:
\begin{equation}
\Theta =  \begin{pmatrix} \Delta_1 & 0 \\ 0 & \Delta_2 \end{pmatrix} 
+\im/2 \begin{pmatrix} \kappa_1 & 0 \\ 0 & \kappa_2 \end{pmatrix}.
\end{equation}

\section{Derivation of the determinant of the coupling matrix \textit{D}}
As shown in equation~(\ref{eq:SI_SdifC}), the contrast between forward and backward transmission coefficients is directly proportional with the determinant of the mode-port coupling matrix $D$ thus breaking the reciprocity demands $\det(D) \neq 0$. Here, we show that this determinant can in general be obtained in terms of the out-coupling losses of the optical modes. 
In fact, the energy conservation relation $D\dg D=K_e$ implies that:
\begin{equation}
|\det(D)|^2=\det (K_e) =\eta_1\kappa_1\eta_2\kappa_2,
\label{eq:SI_detSq}
\end{equation}
where $\eta_1=\frac{\kappa_{e_{1}}}{\kappa_1}$ and $\eta_2=\frac{\kappa_{e_{2}}}{\kappa_2}$ represent the ratio of out-coupling to total losses for each mode. 
On the other hand, the time reversal symmetry requirement of the optical system, i.e., $CD\as=-D$ imposes a condition on the phase of this determinant according to $\det(C) \det(D)\as = \det(D)$.
Multiplying left and right with $\det(D)$ yields $\det(C) |\det(D)|^2=(\det(D))^2$. Equation~(\ref{eq:SI_detSq}) now directly gives
\begin{equation}
\det(D)=\sqrt{\eta_1\eta_2\kappa_1\kappa_2 \det C}.
\end{equation}
Note that in general, C is a unitary matrix with $|\det(C)|=1$. 
In addition, by adjusting the evaluation point at the two ports one can properly choose the phase of $\det(C)$. 
For the system we consider in section E, we have chosen $\det(C)=-1$, which results in $\det(D)=\im\sqrt{\eta_1\eta_2\kappa_1\kappa_2}$. 

According to this relation a necessary condition for breaking the reciprocity is that $\eta_1\neq 0$ and $\eta_2\neq 0$. 
This latter means that both optical modes should be coupled to the ports, or equivalently, the number of independent decay ports ($l$=rank($D$)) should be equal to the number of the actual ports. A possible scenario that violates this condition is the presence of a dark state which is decoupled from the two ports of the system [3]. This is in fact a scenario that would arise when diagonalizing a system that consists of two modes that have equal symmetry with respect to the output channels, through which they would couple at finite rate $\kappa_r$.

\section{Microring resonator system}
Here we explore the microring resonator as a specific case of the general formalism derived in previous sections. In such structure, the direct scattering matrix $C$ can be written as:
\begin{equation}
C=\begin{pmatrix} 0&1\\ 1&0 \end{pmatrix},
\end{equation}
Here, instead of using the clockwise $a_\mathrm{cw}$ and counterclockwise $a_\mathrm{ccw}$ traveling modes, we need to consider the even and odd standing modes, $a_1=(a_\mathrm{cw}+a_\mathrm{ccw})/\sqrt{2}$ and $a_2=(a_\mathrm{cw}-a_\mathrm{ccw})/\im\sqrt{2}$, as our base modes, such that equations~(\ref{eq:SI_timeEvolution}) and~(\ref{eq:SIoutputFields}) are consistent. 
We consider a potentially non-zero frequency splitting $\dd\oo_2-\dd\oo_1$ between the two even and odd modes. 
Therefore, the $\Theta$ matrix involving resonance frequencies can be written as in Eq.~(\ref{eq:SI_Theta}). 
On the other hand, the symmetry/anti-symmetry of the even/odd mode directly implies [1] $d_{11}=d_{21}$ and $d_{12}=-d_{22}$. 
The time reversal symmetry requirement, $CD\as=-D$, therefore leads to $d_{11}\as=-d_{11}$ and $d_{22}\as=d_{22}$. 
While the power conservation relation, $D\dg D = K_e$ leads to $\kappa_{e_{1}}=2|d_{11}|^2, \kappa_{e_{2}}=2|d_{22}|^2$, thus $D$ is obtained as
\begin{equation}
D=\frac{1}{\sqrt{2}} 
\begin{pmatrix}
\im\sqrt{\kappa_{e_{1}}} & -\sqrt{\kappa_{e_{2}}} \\
\im\sqrt{\kappa_{e_{1}}} & \sqrt{\kappa_{e_{2}}} 
\end{pmatrix}.
\label{eq:SI_Dmatrix}
\end{equation}
As a result, for a single drive field with amplitude $\bar{\text{s}}_\text{control}$ incident through port 1 and using $G_1=G_2=G$, the coupling rates $g_1$ and $g_2$ are now given by
\begin{equation}
\begin{pmatrix} g_1\\g_2 \end{pmatrix}
= G\xzpf \begin{pmatrix} \alpha_1\\ \alpha_2 \end{pmatrix}
= G\xzpf 
\begin{pmatrix}
\Soo^{-1}(\omega=0) & 0 \\
0 & \Sot^{-1}(\omega=0)
\end{pmatrix} D^\text{T}
\begin{pmatrix} \bar{\text{s}}_\text{control} \\ 0 \end{pmatrix}
=-\frac{G\xzpf \, \bar{\text{s}}_\text{control}}{\sqrt{2}} 
\begin{pmatrix}
\frac{\sqrt{\eta_1\kappa_1}}{\Db_{1}+\im\kappa_1/2} \\ 
\im \frac{\sqrt{\eta_2\kappa_2}}{\Db_{2}+\im\kappa_2/2} 
\end{pmatrix}.
\end{equation}
Thus for large detuning $|\Db_{1,2}|\gg\kappa_{1,2}$, the optimal phase difference $\Delta\phi=\pi/2$  is automatically satisfied by pumping through a single channel.

Based on Eqs.~(\ref{eq:SI_GenScatMat}) and~(\ref{eq:SI_Dmatrix}), the scattering matrix is written as
\begin{equation}
S= \begin{pmatrix} 0&1\\ 1&0 \end{pmatrix} 
+\frac{\im}{2} 
\begin{pmatrix}
\im\sqrt{\kappa_{e_{1}}} & -\sqrt{\kappa_{e_{2}}} \\
\im\sqrt{\kappa_{e_{1}}} & \sqrt{\kappa_{e_{2}}} 
\end{pmatrix} 
\begin{pmatrix}
\Soo \mp \frac{|g_1|^2}{\Sm^\pm} & \mp \frac{g_1 g_2\as}{\Sm^\pm} \\
\mp \frac{g_1\as g_2}{\Sm^\pm} & \Sot \mp \frac{|g_2|^2}{\Sm^\pm}
\end{pmatrix}^{\!\!-1}
\begin{pmatrix}
\im\sqrt{\kappa_{e_{1}}} &\im\sqrt{\kappa_{e_{1}}} \\
 -\sqrt{\kappa_{e_{2}}} & \sqrt{\kappa_{e_{2}}} 
\end{pmatrix}.
\end{equation}
From here, the forward and backward transmission coefficients are obtained as follows:
\begin{equation}
S_{12}=1-\frac{\im}{2} \frac{\ketwo(\Soo\Sm^\pm \mp |g_1|^2)+\keone(\Sot\Sm^\pm\mp |g_2|^2)\mp\im\sqrt{\keone\ketwo}(g_1 g_2\as-g_1\as g_2)}{\Sm^\pm\Soo\Sot\mp\Sot|g_1|^2\mp\Soo|g_2|^2},
\end{equation}
\begin{equation}
S_{21}=1-\frac{\im}{2} \frac{\ketwo(\Soo\Sm^\pm \mp |g_1|^2)+\keone(\Sot\Sm^\pm\mp |g_2|^2)\pm\im\sqrt{\keone\ketwo}(g_1 g_2\as-g_1\as g_2)}{\Sm^\pm\Soo\Sot\mp\Sot|g_1|^2\mp\Soo|g_2|^2}.
\end{equation}
These latter relations can be rewritten in terms of multi-photon cooperativities of both modes as follows:
\begin{equation}
S_{12}=1-\im\frac{\eta_2\left((\dd_1+\im)(\dd_\pm+\im)\mp\C_1\right)+\eta_1\left((\dd_2+\im)(\dd_\pm+\im)\mp\C_2\right)\mp 2\sqrt{\eta_1\eta_2}\sqrt{\C_1\C_2} \sin(\Delta\phi)}
{(\dd_1+\im)(\dd_2+\im)(\dd_\pm+\im)\mp(\dd_2+\im)\C_1\mp(\dd_1+\im)\C_2},
\end{equation}
\begin{equation}
S_{21}=1-\im\frac{\eta_2\left((\dd_1+\im)(\dd_\pm+\im)\mp\C_1\right)+\eta_1\left((\dd_2+\im)(\dd_\pm+\im)\mp\C_2\right)\pm 2\sqrt{\eta_1\eta_2}\sqrt{\C_1\C_2} \sin(\Delta\phi)}
{(\dd_1+\im)(\dd_2+\im)(\dd_\pm+\im)\mp(\dd_2+\im)\C_1\mp(\dd_1+\im)\C_2}.
\end{equation}

\subsection*{Optical modes with identical loss and coupling}
In order to explore the maximum contrast between the forward and backward transmission coefficients, we consider a resonant probe excitation corresponding to $\oo=\pm\Om$.
Furthermore, the two optical modes exhibit the same amount of losses $(\kappa_{e_{1,2}}=\kappa_e$, $\kappa_{0_{1,2}}=\kappa_0$), and for simplicity we also assume that both modes are pumped with the same internal cavity photon numbers $(|g_{1,2}|=|g|)$ while they can in general exhibit different phases. 
In this case the scattering parameters can be greatly simplified to:
\begin{align}
S_{12}=&1-\eta \frac{2\pm\C (1+\sin(\Delta\phi))} {1\pm\C+\beta^2}, \\
S_{21}=&1-\eta \frac{2\pm\C (1-\sin(\Delta\phi))} {1\pm\C+\beta^2}, \label{eq:SI_S21Simple}
\end{align}
where $\C=2\C_1=2\C_2$ and $\beta=(\pm\Om+\Db_1)/(\kappa/2)=-(\pm\Om+\Db_2)/(\kappa/2)$.
Note that both transmissions are real quantities and their contrast is obtained as
\begin{equation}
S_{21}-S_{12}=2\eta \frac{\pm\C\sin(\Delta\phi)} {1\pm\C+\beta^2},
\end{equation}
This relation clearly shows that maximum contrast can be obtained for $\Delta\phi=\pi/2$, i.e., $g_2=\im g_1$. 
Under the condition of critical coupling $(\eta=\kappa_e/\kappa=1/2)$ the difference between the transmittivities of the forward and backward probes is obtained from
\begin{equation}
\Delta T=|S_{21}|^2 - |S_{12}|^2 = \frac{\C(\C\pm 2\beta^{2})}{(1\pm\C+\beta^{2})^{2}},
\label{eq:SI_DeltaT}
\end{equation}
which is similar to equation~(\ref{eq:DeltaT}) in the main text and can be further simplified to $\Delta T=\C^2/(\C\pm 1)^2$ for the case of degenerate modes.
This latter relation shows that the contrast between forward and backward transmission coefficients is monotonically increasing with cooperativity and asymptotically approaches unity for $\C\rightarrow\infty$. In addition, it is possible to properly adjust a finite $\C$ so that the backward probe can be completely blocked, i.e., $S_{12}=0$. 
From Eq.~(\ref{eq:SI_S21Simple}), for the red detuned regime, such relation is obtained to be:
\begin{equation}
(2\eta-1)(1+\C)=\beta^2
\label{eq:SI_blocking}
\end{equation}
Note that this relation can only be satisfied for a strongly coupled system, i.e., when $\eta>1/2$. For zero mode splitting $(\beta=0)$ equation~(\ref{eq:SI_blocking}) reduces to the condition of critical coupling $\eta=1/2$. In such scenario, independent from the cooperativity, the backward probe can always be fully blocked, while larger cooperativities can increase the forward transmission.

\section{Experimental details}
\subsection*{Experimental setup}
The silica microtoroid (diameter = 41 $\mu$m) is fabricated using techniques as previously reported (see for example reference [4]). A tuneable fibre-coupled external cavity diode laser (New Focus, TLB-6728) is locked to a mechanical sideband of a whispering gallery mode at 1542 nm using a modified Pound-Drever-Hall scheme that can be used independent of the probe beam direction. The probe light is generated using a commercial Double-Parallel Mach-Zehnder Interferometer (DPMZI) (Thorlabs, LN86S-FC) operated in single-sideband carrier-suppressed mode, driven by the output of a VNA at frequency $\Omega$. The resulting probe light has frequency $\opr=\octrl\pm\Omega$. The sign of the frequency shift, as well as the suppression of the carrier (by 50 dB with respect to the generated probe) is controlled by bias voltages applied to the DPMZI. Pump and probe amplitude and polarization are controlled with Variable Optical Attenuators (VOA) and Fibre Polarization Controllers (FPC) (Fig.~\ref{fig:optomechanics_isolation}a). The probe beam propagating in forward or backward direction is recombined with the control beam and their beat on fast (125 MHz) low-noise photo receivers (D1/D2) is analyzed with a VNA. It should be noted that fluctuations of the optical length difference of probe and control paths generate phase fluctuations of the beat analysed by the VNA. To minimize these phase fluctuations on the time scale of the inverse bandwidth (5 kHz)$^{-1}$ of the VNA, the lengths of the paths Laser/C1/D2 and Laser/Switch/C1/D2 are matched, as well as those of the paths Laser/D1 and Laser/Switch/C2/D1 (see Fig.~\ref{fig:optomechanics_isolation}a).    
\subsection*{Measurement procedure and fitting}
Before each measurement the probe power in both propagation directions is balanced using a VOA in one of the probe arms. The polarization of both probe directions is controlled via FPCs, which are tuned separately to optimize the fibre to cavity-mode coupling. To calibrate the transmittance at the probe frequency, a reference measurement is performed with control and probe tuned away from the cavity resonance. Both the reference and measurement are averages of 75 traces of a frequency-swept probe. For each measurement, $|S_{ij}|^2$ is fitted over a wide ω range used to determine $\Db_j$ and $\kappa_j$. Fixing these values, the same equation is fitted to a smaller frequency range surrounding the OMIT peak to yield values for η$\eta_j$ and $|g_j|$. In all fits, $\Om/2\pi$ and $\Gm/2\pi$ are kept fixed at the independently determined values from thermal noise spectra obtained with a spectrum analyzer. The theory curve in figure~\ref{fig:powerDep_Amp}a is obtained using the average value $\eta_j=0.453$ as determined from the four measurements at different control powers. \\

\noindent\textbf{References}\\
\noindent [1] Suh, W., Wang, Z., \& Fan, S. Temporal coupled-mode theory and the presence of non-orthogonal modes in lossless multimode cavities. \textit{IEEE Journal of Quantum Electronics}, \textbf{40}, 1511 (2004). 

\noindent [2] Arfken, G. (1985). Mathematical Methods for Physicists. Academic Press, San Diego.

\noindent [3] Gentry, C. M., \& Popovi\'{c}, M. A. Dark state lasers. \textit{Optics letters}, \textbf{39}, 4136 (2014). 

\noindent [4] A. Schliesser, R. Rivi\`{e}re, G. Anetsberger, O. Arcizet, and T. J. Kippenberg. Resolved sideband cooling of a micromechanical oscillator. \textit{Nature Physics}, \textbf{4}, 415 (2008).


\begin{thebibliography}{31}%
\makeatletter
\providecommand \@ifxundefined [1]{%
 \@ifx{#1\undefined}
}%
\providecommand \@ifnum [1]{%
 \ifnum #1\expandafter \@firstoftwo
 \else \expandafter \@secondoftwo
 \fi
}%
\providecommand \@ifx [1]{%
 \ifx #1\expandafter \@firstoftwo
 \else \expandafter \@secondoftwo
 \fi
}%
\providecommand \natexlab [1]{#1}%
\providecommand \enquote  [1]{``#1''}%
\providecommand \bibnamefont  [1]{#1}%
\providecommand \bibfnamefont [1]{#1}%
\providecommand \citenamefont [1]{#1}%
\providecommand \href@noop [0]{\@secondoftwo}%
\providecommand \href [0]{\begingroup \@sanitize@url \@href}%
\providecommand \@href[1]{\@@startlink{#1}\@@href}%
\providecommand \@@href[1]{\endgroup#1\@@endlink}%
\providecommand \@sanitize@url [0]{\catcode `\\12\catcode `\$12\catcode
  `\&12\catcode `\#12\catcode `\^12\catcode `\_12\catcode `\%12\relax}%
\providecommand \@@startlink[1]{}%
\providecommand \@@endlink[0]{}%
\providecommand \url  [0]{\begingroup\@sanitize@url \@url }%
\providecommand \@url [1]{\endgroup\@href {#1}{\urlprefix }}%
\providecommand \urlprefix  [0]{URL }%
\providecommand \Eprint [0]{\href }%
\providecommand \doibase [0]{http://dx.doi.org/}%
\providecommand \selectlanguage [0]{\@gobble}%
\providecommand \bibinfo  [0]{\@secondoftwo}%
\providecommand \bibfield  [0]{\@secondoftwo}%
\providecommand \translation [1]{[#1]}%
\providecommand \BibitemOpen [0]{}%
\providecommand \bibitemStop [0]{}%
\providecommand \bibitemNoStop [0]{.\EOS\space}%
\providecommand \EOS [0]{\spacefactor3000\relax}%
\providecommand \BibitemShut  [1]{\csname bibitem#1\endcsname}%
\let\auto@bib@innerbib\@empty
\bibitem [{\citenamefont {De{\'{a}}k}\ and\ \citenamefont
  {F{\"{u}}l{\"{o}}p}(2012)}]{Deak2012}%
  \BibitemOpen
  \bibfield  {author} {\bibinfo {author} {\bibfnamefont {L.}~\bibnamefont
  {De{\'{a}}k}}\ and\ \bibinfo {author} {\bibfnamefont {T.}~\bibnamefont
  {F{\"{u}}l{\"{o}}p}},\ }\href {\doibase 10.1016/j.aop.2011.10.013} {\bibfield
   {journal} {\bibinfo  {journal} {Ann. Phys.}\ }\textbf {\bibinfo {volume}
  {327}},\ \bibinfo {pages} {1050} (\bibinfo {year} {2012})}\BibitemShut
  {NoStop}%
\bibitem [{\citenamefont {Haldane}\ and\ \citenamefont
  {Raghu}(2008)}]{Haldane2008}%
  \BibitemOpen
  \bibfield  {author} {\bibinfo {author} {\bibfnamefont {F.~D.~M.}\
  \bibnamefont {Haldane}}\ and\ \bibinfo {author} {\bibfnamefont
  {S.}~\bibnamefont {Raghu}},\ }\href {\doibase 10.1103/PhysRevLett.100.013904}
  {\bibfield  {journal} {\bibinfo  {journal} {Phys. Rev. Lett.}\ }\textbf
  {\bibinfo {volume} {100}},\ \bibinfo {pages} {013904} (\bibinfo {year}
  {2008})}\BibitemShut {NoStop}%
\bibitem [{\citenamefont {Sounas}\ \emph {et~al.}(2013)\citenamefont {Sounas},
  \citenamefont {Caloz},\ and\ \citenamefont {Al{\`{u}}}}]{Sounas2013}%
  \BibitemOpen
  \bibfield  {author} {\bibinfo {author} {\bibfnamefont {D.~L.}\ \bibnamefont
  {Sounas}}, \bibinfo {author} {\bibfnamefont {C.}~\bibnamefont {Caloz}}, \
  and\ \bibinfo {author} {\bibfnamefont {A.}~\bibnamefont {Al{\`{u}}}},\ }\href
  {\doibase 10.1038/ncomms3407} {\bibfield  {journal} {\bibinfo  {journal}
  {Nat. Commun.}\ }\textbf {\bibinfo {volume} {4}},\ \bibinfo {pages} {2407}
  (\bibinfo {year} {2013})}\BibitemShut {NoStop}%
\bibitem [{\citenamefont {Guo}\ \emph {et~al.}(2015)\citenamefont {Guo},
  \citenamefont {Zou}, \citenamefont {Jung},\ and\ \citenamefont
  {Tang}}]{Guo2015}%
  \BibitemOpen
  \bibfield  {author} {\bibinfo {author} {\bibfnamefont {X.}~\bibnamefont
  {Guo}}, \bibinfo {author} {\bibfnamefont {C.-L.}\ \bibnamefont {Zou}},
  \bibinfo {author} {\bibfnamefont {H.}~\bibnamefont {Jung}}, \ and\ \bibinfo
  {author} {\bibfnamefont {H.~X.}\ \bibnamefont {Tang}},\ }\href
  {http://arxiv.org/abs/1511.08112} {\bibfield  {journal} {\bibinfo  {journal}
  {ArXiv}\ } (\bibinfo {year} {2015})},\ \Eprint
  {http://arxiv.org/abs/1511.08112} {arXiv:1511.08112} \BibitemShut {NoStop}%
\bibitem [{\citenamefont {Sayrin}\ \emph {et~al.}(2015)\citenamefont {Sayrin},
  \citenamefont {Junge}, \citenamefont {Mitsch}, \citenamefont {Albrecht},
  \citenamefont {O'Shea}, \citenamefont {Schneeweiss}, \citenamefont {Volz},\
  and\ \citenamefont {Rauschenbeutel}}]{Sayrin2015}%
  \BibitemOpen
  \bibfield  {author} {\bibinfo {author} {\bibfnamefont {C.}~\bibnamefont
  {Sayrin}}, \bibinfo {author} {\bibfnamefont {C.}~\bibnamefont {Junge}},
  \bibinfo {author} {\bibfnamefont {R.}~\bibnamefont {Mitsch}}, \bibinfo
  {author} {\bibfnamefont {B.}~\bibnamefont {Albrecht}}, \bibinfo {author}
  {\bibfnamefont {D.}~\bibnamefont {O'Shea}}, \bibinfo {author} {\bibfnamefont
  {P.}~\bibnamefont {Schneeweiss}}, \bibinfo {author} {\bibfnamefont
  {J.}~\bibnamefont {Volz}}, \ and\ \bibinfo {author} {\bibfnamefont
  {A.}~\bibnamefont {Rauschenbeutel}},\ }\href {\doibase
  10.1103/PhysRevX.5.041036} {\bibfield  {journal} {\bibinfo  {journal} {Phys.
  Rev. X}\ }\textbf {\bibinfo {volume} {5}},\ \bibinfo {pages} {041036}
  (\bibinfo {year} {2015})}\BibitemShut {NoStop}%
\bibitem [{\citenamefont {Kim}\ \emph {et~al.}(2015)\citenamefont {Kim},
  \citenamefont {Kuzyk}, \citenamefont {Han}, \citenamefont {Wang},\ and\
  \citenamefont {Bahl}}]{Kim2015}%
  \BibitemOpen
  \bibfield  {author} {\bibinfo {author} {\bibfnamefont {J.}~\bibnamefont
  {Kim}}, \bibinfo {author} {\bibfnamefont {M.~C.}\ \bibnamefont {Kuzyk}},
  \bibinfo {author} {\bibfnamefont {K.}~\bibnamefont {Han}}, \bibinfo {author}
  {\bibfnamefont {H.}~\bibnamefont {Wang}}, \ and\ \bibinfo {author}
  {\bibfnamefont {G.}~\bibnamefont {Bahl}},\ }\href {\doibase
  10.1038/nphys3236} {\bibfield  {journal} {\bibinfo  {journal} {Nat. Phys.}\
  }\textbf {\bibinfo {volume} {11}},\ \bibinfo {pages} {275} (\bibinfo {year}
  {2015})}\BibitemShut {NoStop}%
\bibitem [{\citenamefont {Dong}\ \emph {et~al.}(2015)\citenamefont {Dong},
  \citenamefont {Shen}, \citenamefont {Zou}, \citenamefont {Zhang},
  \citenamefont {Fu},\ and\ \citenamefont {Guo}}]{Dong2015}%
  \BibitemOpen
  \bibfield  {author} {\bibinfo {author} {\bibfnamefont {C.-H.}\ \bibnamefont
  {Dong}}, \bibinfo {author} {\bibfnamefont {Z.}~\bibnamefont {Shen}}, \bibinfo
  {author} {\bibfnamefont {C.-L.}\ \bibnamefont {Zou}}, \bibinfo {author}
  {\bibfnamefont {Y.-L.}\ \bibnamefont {Zhang}}, \bibinfo {author}
  {\bibfnamefont {W.}~\bibnamefont {Fu}}, \ and\ \bibinfo {author}
  {\bibfnamefont {G.-C.}\ \bibnamefont {Guo}},\ }\href {\doibase
  10.1038/ncomms7193} {\bibfield  {journal} {\bibinfo  {journal} {Nat.
  Commun.}\ }\textbf {\bibinfo {volume} {6}},\ \bibinfo {pages} {6193}
  (\bibinfo {year} {2015})}\BibitemShut {NoStop}%
\bibitem [{\citenamefont {Tzuang}\ \emph {et~al.}(2014)\citenamefont {Tzuang},
  \citenamefont {Fang}, \citenamefont {Nussenzveig}, \citenamefont {Fan},\ and\
  \citenamefont {Lipson}}]{Tzuang2014}%
  \BibitemOpen
  \bibfield  {author} {\bibinfo {author} {\bibfnamefont {L.~D.}\ \bibnamefont
  {Tzuang}}, \bibinfo {author} {\bibfnamefont {K.}~\bibnamefont {Fang}},
  \bibinfo {author} {\bibfnamefont {P.}~\bibnamefont {Nussenzveig}}, \bibinfo
  {author} {\bibfnamefont {S.}~\bibnamefont {Fan}}, \ and\ \bibinfo {author}
  {\bibfnamefont {M.}~\bibnamefont {Lipson}},\ }\href {\doibase
  10.1038/nphoton.2014.177} {\bibfield  {journal} {\bibinfo  {journal} {Nat.
  Photonics}\ }\textbf {\bibinfo {volume} {8}},\ \bibinfo {pages} {701}
  (\bibinfo {year} {2014})}\BibitemShut {NoStop}%
\bibitem [{\citenamefont {Ranzani}\ and\ \citenamefont
  {Aumentado}(2015)}]{Ranzani2015}%
  \BibitemOpen
  \bibfield  {author} {\bibinfo {author} {\bibfnamefont {L.}~\bibnamefont
  {Ranzani}}\ and\ \bibinfo {author} {\bibfnamefont {J.}~\bibnamefont
  {Aumentado}},\ }\href {\doibase 10.1088/1367-2630/17/2/023024} {\bibfield
  {journal} {\bibinfo  {journal} {New J. Phys.}\ }\textbf {\bibinfo {volume}
  {17}},\ \bibinfo {pages} {023024} (\bibinfo {year} {2015})}\BibitemShut
  {NoStop}%
\bibitem [{\citenamefont {Metelmann}\ and\ \citenamefont
  {Clerk}(2015)}]{Metelmann2015}%
  \BibitemOpen
  \bibfield  {author} {\bibinfo {author} {\bibfnamefont {A.}~\bibnamefont
  {Metelmann}}\ and\ \bibinfo {author} {\bibfnamefont {A.~A.}\ \bibnamefont
  {Clerk}},\ }\href {\doibase 10.1103/PhysRevX.5.021025} {\bibfield  {journal}
  {\bibinfo  {journal} {Phys. Rev. X}\ }\textbf {\bibinfo {volume} {5}},\
  \bibinfo {pages} {021025} (\bibinfo {year} {2015})}\BibitemShut {NoStop}%
\bibitem [{\citenamefont {Shoji}\ \emph {et~al.}(2008)\citenamefont {Shoji},
  \citenamefont {Mizumoto}, \citenamefont {Yokoi}, \citenamefont {Hsieh},\ and\
  \citenamefont {Osgood}}]{Shoji2008}%
  \BibitemOpen
  \bibfield  {author} {\bibinfo {author} {\bibfnamefont {Y.}~\bibnamefont
  {Shoji}}, \bibinfo {author} {\bibfnamefont {T.}~\bibnamefont {Mizumoto}},
  \bibinfo {author} {\bibfnamefont {H.}~\bibnamefont {Yokoi}}, \bibinfo
  {author} {\bibfnamefont {I.~W.}\ \bibnamefont {Hsieh}}, \ and\ \bibinfo
  {author} {\bibfnamefont {R.~M.}\ \bibnamefont {Osgood}},\ }\href {\doibase
  10.1063/1.2884855} {\bibfield  {journal} {\bibinfo  {journal} {Appl. Phys.
  Lett.}\ }\textbf {\bibinfo {volume} {92}},\ \bibinfo {pages} {071117}
  (\bibinfo {year} {2008})}\BibitemShut {NoStop}%
\bibitem [{\citenamefont {Fang}\ \emph {et~al.}(2012)\citenamefont {Fang},
  \citenamefont {Yu},\ and\ \citenamefont {Fan}}]{Fang2012a}%
  \BibitemOpen
  \bibfield  {author} {\bibinfo {author} {\bibfnamefont {K.}~\bibnamefont
  {Fang}}, \bibinfo {author} {\bibfnamefont {Z.}~\bibnamefont {Yu}}, \ and\
  \bibinfo {author} {\bibfnamefont {S.}~\bibnamefont {Fan}},\ }\href {\doibase
  10.1103/PhysRevLett.108.153901} {\bibfield  {journal} {\bibinfo  {journal}
  {Phys. Rev. Lett.}\ }\textbf {\bibinfo {volume} {108}},\ \bibinfo {pages}
  {153901} (\bibinfo {year} {2012})}\BibitemShut {NoStop}%
\bibitem [{\citenamefont {Estep}\ \emph {et~al.}(2014)\citenamefont {Estep},
  \citenamefont {Sounas}, \citenamefont {Soric},\ and\ \citenamefont
  {Al{\`{u}}}}]{Estep2014}%
  \BibitemOpen
  \bibfield  {author} {\bibinfo {author} {\bibfnamefont {N.~A.}\ \bibnamefont
  {Estep}}, \bibinfo {author} {\bibfnamefont {D.~L.}\ \bibnamefont {Sounas}},
  \bibinfo {author} {\bibfnamefont {J.}~\bibnamefont {Soric}}, \ and\ \bibinfo
  {author} {\bibfnamefont {A.}~\bibnamefont {Al{\`{u}}}},\ }\href {\doibase
  10.1038/nphys3134} {\bibfield  {journal} {\bibinfo  {journal} {Nat. Phys.}\
  }\textbf {\bibinfo {volume} {10}},\ \bibinfo {pages} {923} (\bibinfo {year}
  {2014})}\BibitemShut {NoStop}%
\bibitem [{\citenamefont {Sounas}\ and\ \citenamefont
  {Al{\`{u}}}(2014)}]{Sounas2014}%
  \BibitemOpen
  \bibfield  {author} {\bibinfo {author} {\bibfnamefont {D.~L.}\ \bibnamefont
  {Sounas}}\ and\ \bibinfo {author} {\bibfnamefont {A.}~\bibnamefont
  {Al{\`{u}}}},\ }\href {\doibase 10.1021/ph400058y} {\bibfield  {journal}
  {\bibinfo  {journal} {ACS Photonics}\ }\textbf {\bibinfo {volume} {1}},\
  \bibinfo {pages} {198} (\bibinfo {year} {2014})}\BibitemShut {NoStop}%
\bibitem [{\citenamefont {Aspelmeyer}\ \emph {et~al.}(2014)\citenamefont
  {Aspelmeyer}, \citenamefont {Kippenberg},\ and\ \citenamefont
  {Marquardt}}]{Aspelmeyer2014}%
  \BibitemOpen
  \bibfield  {author} {\bibinfo {author} {\bibfnamefont {M.}~\bibnamefont
  {Aspelmeyer}}, \bibinfo {author} {\bibfnamefont {T.~J.}\ \bibnamefont
  {Kippenberg}}, \ and\ \bibinfo {author} {\bibfnamefont {F.}~\bibnamefont
  {Marquardt}},\ }\href {\doibase 10.1103/RevModPhys.86.1391} {\bibfield
  {journal} {\bibinfo  {journal} {Rev. Mod. Phys.}\ }\textbf {\bibinfo {volume}
  {86}},\ \bibinfo {pages} {1391} (\bibinfo {year} {2014})}\BibitemShut
  {NoStop}%
\bibitem [{\citenamefont {Massel}\ \emph {et~al.}(2011)\citenamefont {Massel},
  \citenamefont {Heikkil{\"{a}}}, \citenamefont {Pirkkalainen}, \citenamefont
  {Cho}, \citenamefont {Saloniemi}, \citenamefont {Hakonen},\ and\
  \citenamefont {Sillanp{\"{a}}{\"{a}}}}]{Massel2011}%
  \BibitemOpen
  \bibfield  {author} {\bibinfo {author} {\bibfnamefont {F.}~\bibnamefont
  {Massel}}, \bibinfo {author} {\bibfnamefont {T.~T.}\ \bibnamefont
  {Heikkil{\"{a}}}}, \bibinfo {author} {\bibfnamefont {J.-M.}\ \bibnamefont
  {Pirkkalainen}}, \bibinfo {author} {\bibfnamefont {S.~U.}\ \bibnamefont
  {Cho}}, \bibinfo {author} {\bibfnamefont {H.}~\bibnamefont {Saloniemi}},
  \bibinfo {author} {\bibfnamefont {P.~J.}\ \bibnamefont {Hakonen}}, \ and\
  \bibinfo {author} {\bibfnamefont {M.~A.}\ \bibnamefont
  {Sillanp{\"{a}}{\"{a}}}},\ }\href {\doibase 10.1038/nature10628} {\bibfield
  {journal} {\bibinfo  {journal} {Nature}\ }\textbf {\bibinfo {volume} {480}},\
  \bibinfo {pages} {351} (\bibinfo {year} {2011})}\BibitemShut {NoStop}%
\bibitem [{\citenamefont {Hill}\ \emph {et~al.}(2012)\citenamefont {Hill},
  \citenamefont {Safavi-Naeini}, \citenamefont {Chan},\ and\ \citenamefont
  {Painter}}]{Hill2012}%
  \BibitemOpen
  \bibfield  {author} {\bibinfo {author} {\bibfnamefont {J.~T.}\ \bibnamefont
  {Hill}}, \bibinfo {author} {\bibfnamefont {A.~H.}\ \bibnamefont
  {Safavi-Naeini}}, \bibinfo {author} {\bibfnamefont {J.}~\bibnamefont {Chan}},
  \ and\ \bibinfo {author} {\bibfnamefont {O.}~\bibnamefont {Painter}},\ }\href
  {\doibase 10.1038/ncomms2201} {\bibfield  {journal} {\bibinfo  {journal}
  {Nat. Commun.}\ }\textbf {\bibinfo {volume} {3}},\ \bibinfo {pages} {1196}
  (\bibinfo {year} {2012})}\BibitemShut {NoStop}%
\bibitem [{\citenamefont {Lecocq}\ \emph {et~al.}(2016)\citenamefont {Lecocq},
  \citenamefont {Clark}, \citenamefont {Simmonds}, \citenamefont {Aumentado},\
  and\ \citenamefont {Teufel}}]{Lecocq2016}%
  \BibitemOpen
  \bibfield  {author} {\bibinfo {author} {\bibfnamefont {F.}~\bibnamefont
  {Lecocq}}, \bibinfo {author} {\bibfnamefont {J.~B.}\ \bibnamefont {Clark}},
  \bibinfo {author} {\bibfnamefont {R.~W.}\ \bibnamefont {Simmonds}}, \bibinfo
  {author} {\bibfnamefont {J.}~\bibnamefont {Aumentado}}, \ and\ \bibinfo
  {author} {\bibfnamefont {J.~D.}\ \bibnamefont {Teufel}},\ }\href {\doibase
  10.1103/PhysRevLett.116.043601} {\bibfield  {journal} {\bibinfo  {journal}
  {Phys. Rev. Lett.}\ }\textbf {\bibinfo {volume} {116}},\ \bibinfo {pages}
  {043601} (\bibinfo {year} {2016})}\BibitemShut {NoStop}%
\bibitem [{\citenamefont {Weis}\ \emph {et~al.}(2010)\citenamefont {Weis},
  \citenamefont {Rivi{\`{e}}re}, \citenamefont {Del{\'{e}}glise}, \citenamefont
  {Gavartin}, \citenamefont {Arcizet}, \citenamefont {Schliesser},\ and\
  \citenamefont {Kippenberg}}]{Weis2010}%
  \BibitemOpen
  \bibfield  {author} {\bibinfo {author} {\bibfnamefont {S.}~\bibnamefont
  {Weis}}, \bibinfo {author} {\bibfnamefont {R.}~\bibnamefont {Rivi{\`{e}}re}},
  \bibinfo {author} {\bibfnamefont {S.}~\bibnamefont {Del{\'{e}}glise}},
  \bibinfo {author} {\bibfnamefont {E.}~\bibnamefont {Gavartin}}, \bibinfo
  {author} {\bibfnamefont {O.}~\bibnamefont {Arcizet}}, \bibinfo {author}
  {\bibfnamefont {A.}~\bibnamefont {Schliesser}}, \ and\ \bibinfo {author}
  {\bibfnamefont {T.~J.}\ \bibnamefont {Kippenberg}},\ }\href {\doibase
  10.1126/science.1195596} {\bibfield  {journal} {\bibinfo  {journal}
  {Science}\ }\textbf {\bibinfo {volume} {330}},\ \bibinfo {pages} {1520}
  (\bibinfo {year} {2010})}\BibitemShut {NoStop}%
\bibitem [{\citenamefont {Hafezi}\ and\ \citenamefont
  {Rabl}(2012)}]{Hafezi2012}%
  \BibitemOpen
  \bibfield  {author} {\bibinfo {author} {\bibfnamefont {M.}~\bibnamefont
  {Hafezi}}\ and\ \bibinfo {author} {\bibfnamefont {P.}~\bibnamefont {Rabl}},\
  }\href {\doibase 10.1364/OE.20.007672} {\bibfield  {journal} {\bibinfo
  {journal} {Opt. Express}\ }\textbf {\bibinfo {volume} {20}},\ \bibinfo
  {pages} {7672} (\bibinfo {year} {2012})}\BibitemShut {NoStop}%
\bibitem [{\citenamefont {Shen}\ \emph {et~al.}(2016)\citenamefont {Shen},
  \citenamefont {Zhang}, \citenamefont {Chen}, \citenamefont {Zou},
  \citenamefont {Xiao}, \citenamefont {Zou}, \citenamefont {Sun}, \citenamefont
  {Guo},\ and\ \citenamefont {Dong}}]{Shen2016}%
  \BibitemOpen
  \bibfield  {author} {\bibinfo {author} {\bibfnamefont {Z.}~\bibnamefont
  {Shen}}, \bibinfo {author} {\bibfnamefont {Y.-L.}\ \bibnamefont {Zhang}},
  \bibinfo {author} {\bibfnamefont {Y.}~\bibnamefont {Chen}}, \bibinfo {author}
  {\bibfnamefont {C.-L.}\ \bibnamefont {Zou}}, \bibinfo {author} {\bibfnamefont
  {Y.-F.}\ \bibnamefont {Xiao}}, \bibinfo {author} {\bibfnamefont {X.-B.}\
  \bibnamefont {Zou}}, \bibinfo {author} {\bibfnamefont {F.-W.}\ \bibnamefont
  {Sun}}, \bibinfo {author} {\bibfnamefont {G.-C.}\ \bibnamefont {Guo}}, \ and\
  \bibinfo {author} {\bibfnamefont {C.-H.}\ \bibnamefont {Dong}},\ }\href
  {http://arxiv.org/abs/1604.02297} {\bibfield  {journal} {\bibinfo  {journal}
  {ArXiv}\ } (\bibinfo {year} {2016})},\ \Eprint
  {http://arxiv.org/abs/1604.02297} {arXiv:1604.02297} \BibitemShut {NoStop}%
\bibitem [{\citenamefont {Habraken}\ \emph {et~al.}(2012)\citenamefont
  {Habraken}, \citenamefont {Stannigel}, \citenamefont {Lukin}, \citenamefont
  {Zoller},\ and\ \citenamefont {Rabl}}]{Habraken2012}%
  \BibitemOpen
  \bibfield  {author} {\bibinfo {author} {\bibfnamefont {S.~J.~M.}\
  \bibnamefont {Habraken}}, \bibinfo {author} {\bibfnamefont {K.}~\bibnamefont
  {Stannigel}}, \bibinfo {author} {\bibfnamefont {M.~D.}\ \bibnamefont
  {Lukin}}, \bibinfo {author} {\bibfnamefont {P.}~\bibnamefont {Zoller}}, \
  and\ \bibinfo {author} {\bibfnamefont {P.}~\bibnamefont {Rabl}},\ }\href
  {\doibase 10.1088/1367-2630/14/11/115004} {\bibfield  {journal} {\bibinfo
  {journal} {New J. Phys.}\ }\textbf {\bibinfo {volume} {14}},\ \bibinfo
  {pages} {115004} (\bibinfo {year} {2012})}\BibitemShut {NoStop}%
\bibitem [{\citenamefont {Xu}\ and\ \citenamefont {Li}(2015)}]{Xu2015}%
  \BibitemOpen
  \bibfield  {author} {\bibinfo {author} {\bibfnamefont {X.-W.}\ \bibnamefont
  {Xu}}\ and\ \bibinfo {author} {\bibfnamefont {Y.}~\bibnamefont {Li}},\ }\href
  {\doibase 10.1103/PhysRevA.91.053854} {\bibfield  {journal} {\bibinfo
  {journal} {Phys. Rev. A}\ }\textbf {\bibinfo {volume} {91}},\ \bibinfo
  {pages} {053854} (\bibinfo {year} {2015})}\BibitemShut {NoStop}%
\bibitem [{\citenamefont {Peano}\ \emph {et~al.}(2015)\citenamefont {Peano},
  \citenamefont {Brendel}, \citenamefont {Schmidt},\ and\ \citenamefont
  {Marquardt}}]{Peano2015}%
  \BibitemOpen
  \bibfield  {author} {\bibinfo {author} {\bibfnamefont {V.}~\bibnamefont
  {Peano}}, \bibinfo {author} {\bibfnamefont {C.}~\bibnamefont {Brendel}},
  \bibinfo {author} {\bibfnamefont {M.}~\bibnamefont {Schmidt}}, \ and\
  \bibinfo {author} {\bibfnamefont {F.}~\bibnamefont {Marquardt}},\ }\href
  {\doibase 10.1103/PhysRevX.5.031011} {\bibfield  {journal} {\bibinfo
  {journal} {Phys. Rev. X}\ }\textbf {\bibinfo {volume} {5}},\ \bibinfo {pages}
  {031011} (\bibinfo {year} {2015})}\BibitemShut {NoStop}%
\bibitem [{\citenamefont {Schmidt}\ \emph {et~al.}(2015)\citenamefont
  {Schmidt}, \citenamefont {Kessler}, \citenamefont {Peano}, \citenamefont
  {Painter},\ and\ \citenamefont {Marquardt}}]{Schmidt2015}%
  \BibitemOpen
  \bibfield  {author} {\bibinfo {author} {\bibfnamefont {M.}~\bibnamefont
  {Schmidt}}, \bibinfo {author} {\bibfnamefont {S.}~\bibnamefont {Kessler}},
  \bibinfo {author} {\bibfnamefont {V.}~\bibnamefont {Peano}}, \bibinfo
  {author} {\bibfnamefont {O.}~\bibnamefont {Painter}}, \ and\ \bibinfo
  {author} {\bibfnamefont {F.}~\bibnamefont {Marquardt}},\ }\href {\doibase
  10.1364/OPTICA.2.000635} {\bibfield  {journal} {\bibinfo  {journal} {Optica}\
  }\textbf {\bibinfo {volume} {2}},\ \bibinfo {pages} {635} (\bibinfo {year}
  {2015})}\BibitemShut {NoStop}%
\bibitem [{\citenamefont {Schliesser}\ \emph {et~al.}(2008)\citenamefont
  {Schliesser}, \citenamefont {Rivi{\`{e}}re}, \citenamefont {Anetsberger},
  \citenamefont {Arcizet},\ and\ \citenamefont {Kippenberg}}]{Schliesser2008}%
  \BibitemOpen
  \bibfield  {author} {\bibinfo {author} {\bibfnamefont {A.}~\bibnamefont
  {Schliesser}}, \bibinfo {author} {\bibfnamefont {R.}~\bibnamefont
  {Rivi{\`{e}}re}}, \bibinfo {author} {\bibfnamefont {G.}~\bibnamefont
  {Anetsberger}}, \bibinfo {author} {\bibfnamefont {O.}~\bibnamefont
  {Arcizet}}, \ and\ \bibinfo {author} {\bibfnamefont {T.~J.}\ \bibnamefont
  {Kippenberg}},\ }\href {\doibase 10.1038/nphys939} {\bibfield  {journal}
  {\bibinfo  {journal} {Nat. Phys.}\ }\textbf {\bibinfo {volume} {4}},\
  \bibinfo {pages} {415} (\bibinfo {year} {2008})}\BibitemShut {NoStop}%
\bibitem [{\citenamefont {Yu}\ and\ \citenamefont {Fan}(2009)}]{Yu2009}%
  \BibitemOpen
  \bibfield  {author} {\bibinfo {author} {\bibfnamefont {Z.}~\bibnamefont
  {Yu}}\ and\ \bibinfo {author} {\bibfnamefont {S.}~\bibnamefont {Fan}},\
  }\href {\doibase 10.1063/1.3127531} {\bibfield  {journal} {\bibinfo
  {journal} {Nat. Photonics}\ }\textbf {\bibinfo {volume} {3}},\ \bibinfo
  {pages} {91} (\bibinfo {year} {2009})}\BibitemShut {NoStop}%
\bibitem [{\citenamefont {Suh}\ \emph {et~al.}(2004)\citenamefont {Suh},
  \citenamefont {Wang},\ and\ \citenamefont {Fan}}]{Suh2004}%
  \BibitemOpen
  \bibfield  {author} {\bibinfo {author} {\bibfnamefont {W.}~\bibnamefont
  {Suh}}, \bibinfo {author} {\bibfnamefont {Z.}~\bibnamefont {Wang}}, \ and\
  \bibinfo {author} {\bibfnamefont {S.}~\bibnamefont {Fan}},\ }\href {\doibase
  10.1109/JQE.2004.834773} {\bibfield  {journal} {\bibinfo  {journal} {IEEE J.
  Quant. Electron.}\ }\textbf {\bibinfo {volume} {40}},\ \bibinfo {pages}
  {1511} (\bibinfo {year} {2004})}\BibitemShut {NoStop}%
\bibitem [{\citenamefont {Verhagen}\ \emph {et~al.}(2012)\citenamefont
  {Verhagen}, \citenamefont {Del{\'{e}}glise}, \citenamefont {Weis},
  \citenamefont {Schliesser},\ and\ \citenamefont {Kippenberg}}]{Verhagen2012}%
  \BibitemOpen
  \bibfield  {author} {\bibinfo {author} {\bibfnamefont {E.}~\bibnamefont
  {Verhagen}}, \bibinfo {author} {\bibfnamefont {S.}~\bibnamefont
  {Del{\'{e}}glise}}, \bibinfo {author} {\bibfnamefont {S.}~\bibnamefont
  {Weis}}, \bibinfo {author} {\bibfnamefont {A.}~\bibnamefont {Schliesser}}, \
  and\ \bibinfo {author} {\bibfnamefont {T.~J.}\ \bibnamefont {Kippenberg}},\
  }\href {\doibase 10.1038/nature10787} {\bibfield  {journal} {\bibinfo
  {journal} {Nature}\ }\textbf {\bibinfo {volume} {482}},\ \bibinfo {pages}
  {63} (\bibinfo {year} {2012})}\BibitemShut {NoStop}%
\bibitem [{\citenamefont {Manipatruni}\ \emph {et~al.}(2009)\citenamefont
  {Manipatruni}, \citenamefont {Robinson},\ and\ \citenamefont
  {Lipson}}]{Manipatruni2009}%
  \BibitemOpen
  \bibfield  {author} {\bibinfo {author} {\bibfnamefont {S.}~\bibnamefont
  {Manipatruni}}, \bibinfo {author} {\bibfnamefont {J.~T.}\ \bibnamefont
  {Robinson}}, \ and\ \bibinfo {author} {\bibfnamefont {M.}~\bibnamefont
  {Lipson}},\ }\href {\doibase 10.1103/PhysRevLett.102.213903} {\bibfield
  {journal} {\bibinfo  {journal} {Phys. Rev. Lett.}\ }\textbf {\bibinfo
  {volume} {102}},\ \bibinfo {pages} {213903} (\bibinfo {year}
  {2009})}\BibitemShut {NoStop}%
\bibitem [{\citenamefont {Fan}\ \emph {et~al.}(2012)\citenamefont {Fan},
  \citenamefont {Wang}, \citenamefont {Varghese}, \citenamefont {Shen},
  \citenamefont {Niu}, \citenamefont {Xuan}, \citenamefont {Weiner},\ and\
  \citenamefont {Qi}}]{Fan2012}%
  \BibitemOpen
  \bibfield  {author} {\bibinfo {author} {\bibfnamefont {L.}~\bibnamefont
  {Fan}}, \bibinfo {author} {\bibfnamefont {J.}~\bibnamefont {Wang}}, \bibinfo
  {author} {\bibfnamefont {L.~T.}\ \bibnamefont {Varghese}}, \bibinfo {author}
  {\bibfnamefont {H.}~\bibnamefont {Shen}}, \bibinfo {author} {\bibfnamefont
  {B.}~\bibnamefont {Niu}}, \bibinfo {author} {\bibfnamefont {Y.}~\bibnamefont
  {Xuan}}, \bibinfo {author} {\bibfnamefont {A.~M.}\ \bibnamefont {Weiner}}, \
  and\ \bibinfo {author} {\bibfnamefont {M.}~\bibnamefont {Qi}},\ }\href
  {\doibase 10.1126/science.1214383} {\bibfield  {journal} {\bibinfo  {journal}
  {Science}\ }\textbf {\bibinfo {volume} {335}},\ \bibinfo {pages} {447}
  (\bibinfo {year} {2012})}\BibitemShut {NoStop}%
\end{thebibliography}
\end{document}